\documentclass[10pt,a4paper]{article}
\usepackage{jheppub_kim}
\usepackage{pdflscape}
\usepackage{amsmath}
\usepackage{amssymb}
\usepackage{dcolumn}
\usepackage{bm}
\usepackage{color}
\usepackage{epsfig}
\usepackage{amsfonts}
\usepackage{graphicx}
\usepackage{subfigure}
\usepackage{dcolumn}

\begin{document}
\title{Five-Dimensional Yang-Mills Black Holes in Massive Gravity's Rainbow}
\author[c]{Houcine Aounallah,}
\author[a]{Behnam Pourhassan,}
\author[b]{Seyed Hossein Hendi,}
\author[d]{Mir Faizal}

\affiliation[a] {School of Physics, Damghan University, Damghan,
3671641167, Iran.} \affiliation[a] {Canadian Quantum Research
Center 204-3002 32 Ave Vernon, BC V1T 2L7 Canada.} \affiliation[b]
{Physics Department and Biruni Observatory, College of Sciences,
Shiraz University, Shiraz 71454, Iran.} \affiliation[b] {Canadian
Quantum Research Center 204-3002 32 Ave Vernon, BC V1T 2L7
Canada.} \affiliation[c] {Department of Science and Technology.
Larbi Tebessi University, 12000 Tebessa, Algeria.} \affiliation[d]
{Department of Physics and Astronomy, University of Lethbridge,
Lethbridge, Alberta, T1K 3M4, Canada.} \affiliation[d] {Irving K.
Barber School of Arts and Sciences, University of British
Columbia, Kelowna, British Columbia, V1V 1V7, Canada.}
\affiliation[d] {Canadian Quantum Research Center 204-3002 32 Ave
Vernon, BC V1T 2L7 Canada.}

\emailAdd{houcine.aounallah@univ-tebessa.dz} %
\emailAdd{b.pourhassan@du.ac.ir}
\emailAdd{hendi@shirazu.ac.ir} %
\emailAdd{mirfaizalmir@gmail.com}

\abstract{ In this paper, we will analyze a five-dimensional
Yang-Mills black hole solution in massive gravity's rainbow. We
will also investigate the flow of such a solution with scale.
Then, we will discuss the scale dependence of the thermodynamics
for this black hole. In addition, we study the criticality in the
extended phase space by treating  the cosmological constant as the
thermodynamics pressure of this black hole solution. Moreover, we
will use the partition function for this solution to obtaining
corrections to the thermodynamics of this system and examine their
key role on the behavior of corrected solutions.}


\maketitle

\section{Introduction}

The Yang-Mills black hole solutions have also been motivated by the bosonic
part of the low-energy heterotic string action \cite{ymph18, ymph17}. This
is done by considering the low-energy heterotic string action to leading
order after it has been compactified to four dimensions. This compactified
action is then used to obtain a static and spherically symmetric Yang-Mills
black hole solution. This has motivated the construction of interesting
Yang-Mills black hole solutions \cite{ym1212, ym1414}. A static spherically
symmetric Yang-Mills black hole solution has been studied both numerically
and analytically, and it was observed that such a solution is unstable under
linearized perturbations \cite{ymph10}. The phase transition for Yang-Mills
black holes has been studied using entanglement entropy and two-point
correlation functions \cite{ymph12, ymph14}.

It is possible for gravitons to become massive in string theory due to
scalar fields acquiring a vacuum expectation values \cite{stma, stma10}. In
fact, motivated by string theory, the corrections to a braneworld model
(with warped AdS spacetime) from graviton mass have been discussed \cite%
{stma12}. So, it is possible for the gravitons to have a small mass \cite%
{deRham10,deRham11,deRham14,Hassan12}, and this gravitons mass is constraint
from LIGO collaboration to $m_g<1.2 \times 10$ $ev/c^2$ \cite{deRham14,LIGO}%
. As it is possible for gravitons to have any mass below this bound, such a
small mass would produce an IR correction to general relativity and have
important astrophysical consequences. In fact, it has been suggested that
such massive gravitons could produce an effective cosmological constant, and
cause an accelerated expansion of the universe \cite{Leon,Akrami13,Akrami15}%
. Thus, it is very important to study the modification to general relativity
from small graviton mass. It is possible to add a mass term in the form of
Fierz-Pauli term to obtain massive gravity \cite{Fierz1939}. However, due to
the vDVZ discontinuity, this theory is not well defined in the zero mass
limit \cite{Dam1970,Zakharov70,Nieuwenhuizen1973}.

This problem can be resolved in the Vainshtein mechanism, by using the
Stueckelberg trick \cite{Vainshtein,Hinterbichler}. This mechanism produces
non-linear corrections terms, which in turn produce ghost states \cite{BD}.
However, these ghost states can be removed in the dRGT gravity \cite%
{deRham10}. As dRGT gravity is an important ghost free IR modification to
general relativity (without the vDVZ discontinuity), several black hole
solutions have been constructed in it \cite%
{Koyama11,Nieuwenhuizen11,Gruzinov11,Berezhiani11,deRhamPLB12,Hendi2016,Suchant16,Do93,Do94,Hendi2017}%
. It has also been demonstrated that the small graviton mass can produce
interesting modifications to the behavior of such black hole solutions. In
fact, it has been observed that the thermodynamics of black holes gets
non-trivial modifications due to the small graviton mass \cite%
{HendiPRDMann,DehghaniEPJC,DehghaniCQG,DehghaniPRD}. Thus, it is interesting
to analyze black hole thermodynamics for different black holes using such a
small graviton mass. So, in this paper, we will analyze the modifications to
a five-dimensional Yang-Mills black hole in massive gravity \cite{action1,
action2, action3, action4}.

It is possible to use extended phase space to analyze an AdS black hole
solution \cite{exten1, exten2}. In such an extended phase space the
cosmological constant is identified with the thermodynamic pressure of the
black hole solution. Furthermore, the thermodynamic volume is the conjugate
variable to this thermodynamic pressure. Thus, the thermodynamic volume can
be obtained from the cosmological constant of an AdS black hole solution.
The Van der Waals like behavior an AdS black hole solution has also been
studied using this extended phase space \cite{Dolan11}. It is possible to
analyze the triplet point for an AdS black hole solution \cite%
{Altamirano14,Wei14,Hennigar15}. In fact, the reentrant phase transitions
for an AdS black hole solution has also been studied using this formalism
\cite{Gunasekaran12,Frassino14,HennigarEntropy}. It is interesting to
analyze critical behavior for black hole solutions in massive gravity, and
the graviton mass can produce new non-trivial phase transitions \cite%
{massive2, massive4}.

It may be noted that it is possible to analyze the thermal corrections to
black hole solutions \cite{T1}. This can be done for an AdS black hole as it
is dual to conformal field theory, and so its microstates can be analyzed
using a conformal field theory \cite{T2}. Thus, using the partition function
for such micro states, it would be possible to analyze the corrections to
the thermodynamics of black holes. The leading order corrections to the
entropy of black holes is a logarithmic correction, it is possible to
analyze the effects of this correction on various other thermodynamic
quantities \cite{EPL, EPJC1}. It has been argued that these thermal
fluctuations can have important consequences for the stability of black hole
solutions \cite{PRD1, NPB}. As the black hole evaporates due to Hawking
radiation, these corrections cannot be neglected. Thus, it is important to
analyze the effects of such corrections for Yang-Mills black holes. It may
be noted that these thermal fluctuations in the thermodynamics of a black
hole can be related to quantum fluctuations using the Jacobson formalism
\cite{thermal1, thermal2}.

As string theory can be viewed as a two-dimensional conformal field theory.
The target space metric can be regarded as a matrix of coupling constants,
and these would flow due to the renormalization group flow \cite{rgflow,
rgflow12}. Thus, the target space geometry would flow with scale, and this
flow would depend on the energy of the probe. This consideration has
motivated gravity's rainbow, where the spacetime geometry depends on the
energy of the probe \cite{AK, ra12, ra14, ra18}. It is known that the
energy-dependence of such geometry can produce important modifications to
the thermodynamics of black holes \cite{faiz1, faiz2}. In fact, it has been
observed that such an energy-dependence can have important consequences for
the detection of mini black holes at the LHC \cite{LHC}. We will use such an
energy-dependent metric of gravity's rainbow for analyzing these Yang-Mills
black hole solutions, as such solutions can be motivated from the bosonic
part of the low-energy heterotic string action \cite{ymph18, ymph17}. In
this formalism, the geometry of the Yang-Mills black hole depends on the
energy of the probe. As the particle emitted in the Hawking radiation can
act as a probe for the geometry of a black hole, it is this energy that
deforms the geometry of Yang-Mills black holes. So, in this paper, will
analyze the scale dependence of the geometry of the Yang-Mills black hole in
massive gravity using different rainbow functions.\newline
The paper is organized as follows. In the next section, we review Yang-Mills
black hole solution in massive gravity together horizon structure analysis.
In section \ref{sec3} we study the thermodynamics of the three separated
models. Criticality in the extended phase space discussed in section \ref%
{sec44}. Then, in section \ref{sec4} we consider the effect of thermal
fluctuations and study corrected thermodynamics. Finally, in section \ref%
{sec6} we give conclusion.

\label{sec1} %%%%%%%%%%%%%%%%%%%%%%%%%%%%%%%%%%

%%%%%%%%%%%%%%%%%%%%%%%%%%%%%%%%%%%%%%%%%%

\section{Yang-Mills Black Hole}

\label{sec2} %%%%%%%%%%%%%%%%%%%%%%%%%%%%%%%%%%%%%%%%%%

In this section, we will analyze the Yang-Mills black hole solution in
massive gravity, and its flow with scale. The action of five-dimensional
massive gravity with negative cosmological constant coupled to Yang-Mills
theory can be written as \cite{action1, action2, action3, action4},
\begin{equation*}
S=\int d^{5}x\sqrt{-g}\left( R-2\Lambda -\gamma _{ab}F_{\mu \nu }^{\left(
a\right) }F^{\left( b\right) \mu \nu }+m^{2}\sum c_{i}U_{i}\left( g,f\right)
\right) ,
\end{equation*}%
where $R$ is the Ricci scalar, $m$ is the mass term of massive gravity, $%
\Lambda =-\frac{6}{l^{2}}$ is the cosmological constant, ($l$ denotes the
AdS radius) and $F_{\mu \nu }^{\left( a\right) }$ is the $SO\left(
5,1\right) $. This Yang-Mills gauge field tensor is given by
\begin{equation*}
F_{\mu \nu }^{\left( a\right) }=\partial _{\mu }A_{\nu }^{\left( a\right)
}-\partial _{\nu }A_{\mu }^{\left( a\right) }+\frac{1}{2e}f_{\left( b\right)
\left( c\right) }^{\left( a\right) }A_{\mu }^{\left( b\right) }A_{\nu
}^{\left( c\right) },
\end{equation*}%
where $e$ is coupling constant of the Yang-Mills theory. Also, $c_{i}$ are
constants and $U_{i}\left( g,f\right) $ are symmetric polynomials of the
eigenvalues of the following $5\times 5$ matrix \cite{action5, action6},
\begin{equation}
K_{\nu }^{\mu }=\sqrt{g^{\mu \alpha }f_{\alpha \nu }},
\end{equation}%
where $f_{\alpha \nu }$ can be expressed as $f_{\mu \nu }=dia\left( 0,0,%
\frac{c_{0}^{2}h_{ij}}{g^{2}\left( \varepsilon \right) }\right) $. Here we
have introduced $f\left( \varepsilon \right) $ and $g\left( \varepsilon
\right) $ as the rainbow functions, which depend on the relative energy $%
\varepsilon =\frac{E}{E_{p}}$, where $E$ is the energy of the particle
emitted in the Hawking radiation, and $E_{p}$ is Planck energy \cite{AK,
ra12, ra14, ra18}. This is because string theory is a two-dimensional
conformal field theory, with the target space metric as a matrix of coupling
constants for that conformal field theory. So, this matrix of coupling
constants is expected to flow due to the renormalization group flow \cite%
{rgflow, rgflow12}. This would make the geometry of spacetime depend on the
ratio $\mu /\mu _{p}$, where $\mu $ is the scale at which the theory is
being probed and $\mu _{p}$ is the Planck scale. Now, this ratio would be
proportional to $\varepsilon =\frac{E}{E_{p}}$, and so the geometry of
spacetime should be a function of this ratio. So, the renormalization group
flow would make the target space geometry depend on the scale at which it is
being probed, and this in turn would depend on the energy of the probe. This
energy-dependence of the geometry can be analyzed using these rainbow
function \cite{AK, ra12, ra14, ra18}. In fact, as the Yang-Mills black hole
solutions can be motivated from the bosonic part of the low-energy heterotic
string action \cite{ymph18, ymph17}, we will use analyze the effect of such
a flow of geometry of the Yang-Mills black hole solution. Now the black
holes with AdS asymptote in the massive gravity's rainbow can be described
by the following energy-dependent metric
\begin{equation*}
ds^{2}=-\frac{\psi \left( r\right) }{f^{2}\left( \varepsilon \right) }dt^{2}+%
\frac{1}{g^{2}\left( \varepsilon \right) }\left[ \frac{dr^{2}}{\psi \left(
r\right) }+r^{2}d\Omega _{k}^{2}\right] ,
\end{equation*}%
where $\psi $ is an unknown function which will be determined by field
equations, and%
\begin{equation}
d\Omega _{k}^{2}=\left\{
\begin{array}{cc}
d\theta _{1}^{2}+\sin ^{2}\theta _{1}d\theta _{2}^{2}+\sin ^{2}\theta
_{1}\sin ^{2}\theta _{2}d\theta _{3}^{2} & k=1 \\
d\theta _{1}^{2}+d\theta _{2}^{2}+d\theta _{3}^{2} & k=0 \\
d\theta _{1}^{2}+\sinh ^{2}\theta _{1}d\theta _{2}^{2}+\sinh ^{2}\theta
_{1}\sin ^{2}\theta _{2}d\theta _{3}^{2} & k=-1%
\end{array}%
\right. ,
\end{equation}%
where $k=1$, $0$ and $-1$ represent spherical, flat and hyperbolic horizon
of possible black holes, respectively. Hereafter, we indicate $\omega _{k}$
as the volume of boundary $t=cte$ and $r=cte$ of the metric. Using $\left[ K%
\right] =Tra(K)=K_{\mu }^{\mu }$, one can obtain
\begin{eqnarray}
U_{1} &=&\left[ K\right] =\frac{3c_{0}}{r},  \notag  \label{eq04} \\
U_{2} &=&\left[ K\right] ^{2}-\left[ K^{2}\right] =\frac{6c_{0}^{2}}{r^{2}},
\notag \\
U_{3} &=&\left[ K\right] ^{3}-3\left[ K\right] \left[ K^{2}\right] +2\left[
K^{3}\right] =\frac{6c_{0}^{3}}{r^{3}},  \notag \\
U_{4} &=&\left[ K\right] ^{4}-6\left[ K^{2}\right] \left[ K\right] ^{2}+8%
\left[ K^{3}\right] \left[ K\right] +3\left[ K^{2}\right] ^{2}-6\left[ K^{4}%
\right] .
\end{eqnarray}%
By using the variational principle, one can obtain the following field
equations
\begin{eqnarray}
R_{\mu \nu }+\left( \Lambda -\frac{R}{2}\right) g_{\mu \nu }-m^{2}\chi _{\mu
\nu } &=&8\pi T_{\mu \nu }. \\
F_{;\nu }^{\left( a\right) \mu \nu } &=&J^{\left( a\right) \mu },
\end{eqnarray}%
where
\begin{eqnarray}
T_{\mu \nu } &=&\frac{1}{4\pi }\gamma _{ab}\left( F_{\mu }^{\left( a\right)
\lambda }F_{\nu \lambda }^{\left( b\right) }-\frac{1}{4}F^{\left( a\right)
\lambda \sigma }F_{\lambda \sigma }^{\left( b\right) }g_{\mu \nu }\right) ,
\\
J^{\left( a\right) \nu } &=&\frac{1}{e}f_{\left( b\right) \left( c\right)
}^{\left( a\right) }A_{\mu }^{\left( b\right) }F^{\left( c\right) \mu \nu }.
\end{eqnarray}%
Furthermore, we also have
\begin{eqnarray}
&&\chi _{\mu \nu }=\frac{c_{1}}{2}(U_{1}g_{\mu \nu }-K_{\mu \nu })+\frac{%
c_{2}}{2}(U_{2}g_{\mu \nu }-2U_{1}K_{\mu \nu }+2K_{\mu \nu }^{2})  \notag
\label{eq12} \\
&+&\frac{c_{3}}{2}(U_{3}g_{\mu \nu }-3U_{2}K_{\mu \nu }+6U_{1}K_{\mu \nu
}^{2}-6K_{\mu \nu }^{3})  \notag \\
&+&\frac{c_{4}}{2}(U_{4}g_{\mu \nu }-4U_{3}K_{\mu \nu }+12U_{2}K_{\mu \nu
}^{2}-24U_{1}K_{\mu \nu }^{3}+24K_{\mu \nu }^{4})
\end{eqnarray}%
By using the value of the Yang-Mills field $F=\gamma _{ab}F^{\left( a\right)
\mu \nu }F_{\mu \nu }^{\left( b\right) }$, which is $F=\frac{6e^{2}}{r^{4}}$
in five-dimensions, we can obtain $rr-$component of the field equation as
\begin{equation}
R_{11}+\left( \Lambda -\frac{R}{2}\right) g_{11}-m^{2}\chi _{11}=8\pi T_{11}
\label{FE11}
\end{equation}%
where
\begin{equation*}
R_{11}=-\frac{1}{2}\frac{\psi ^{^{\prime \prime }}\left( r\right) }{\psi
\left( r\right) }-\frac{3}{2}\frac{\psi ^{^{\prime }}\left( r\right) }{r\psi
\left( r\right) },
\end{equation*}%
\begin{equation*}
R=-g^{2}\left( \varepsilon \right) \left( \psi ^{^{\prime \prime }}\left(
r\right) +6\frac{\psi ^{^{\prime }}\left( r\right) }{r}+6\frac{\psi \left(
r\right) }{r^{2}}-\frac{6k}{r^{2}}\right) ,
\end{equation*}%
\begin{equation*}
\chi _{11}=\frac{3}{2g^{2}\left( \varepsilon \right) \psi \left( r\right) }%
\left( \frac{c_{0}c_{1}}{r}+\frac{2c_{0}^{2}c_{2}}{r^{2}}+\frac{%
2c_{0}^{3}c_{3}}{r^{3}}\right) ,
\end{equation*}%
\begin{equation*}
T_{11}=\frac{-3e^{2}}{8\pi g^{2}\left( \varepsilon \right) \psi \left(
r\right) r^{4}}.
\end{equation*}

\begin{figure}[tbp]
\begin{center}
$%
\begin{array}{cccc}
\includegraphics[width=65 mm]{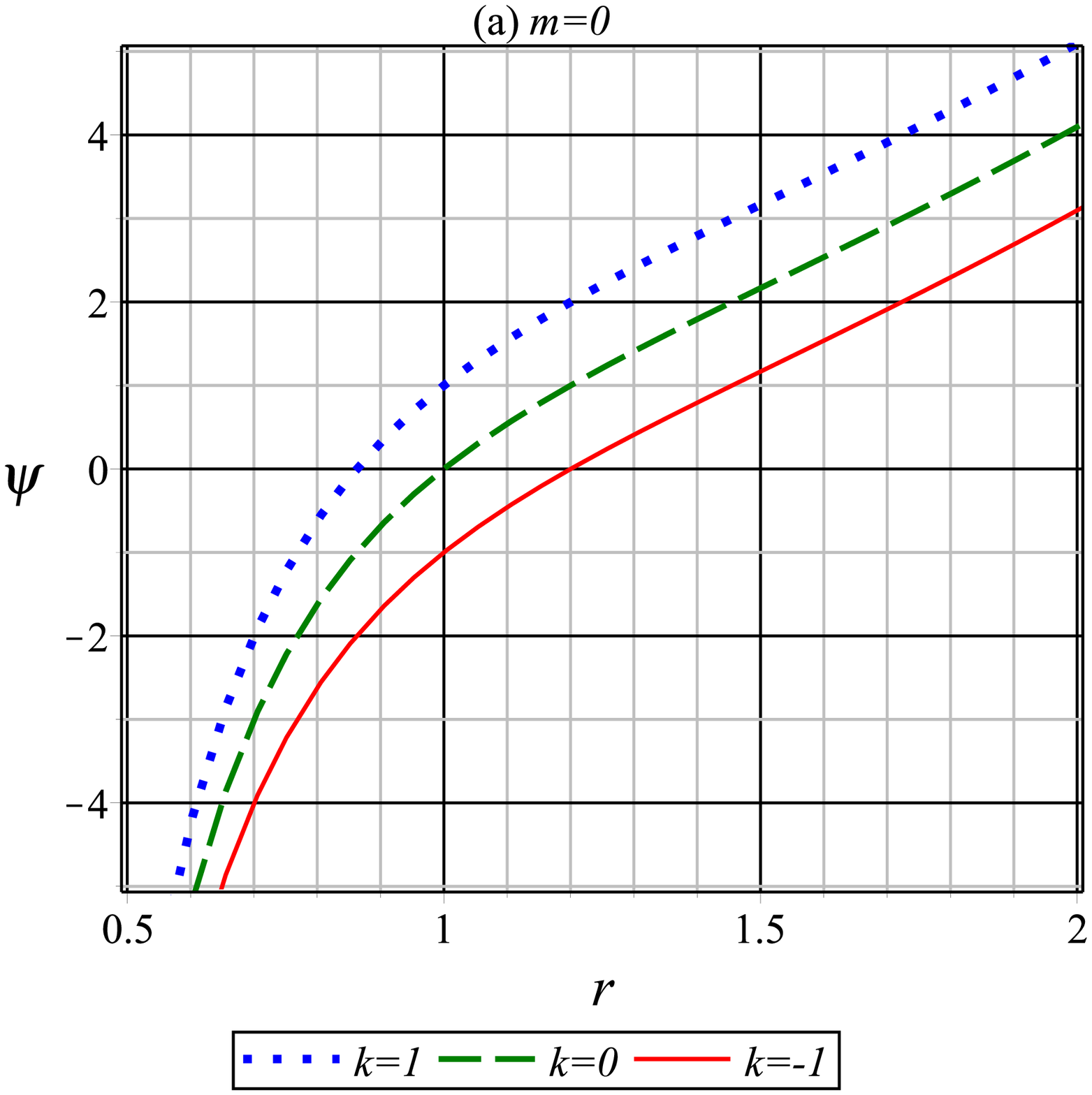}\includegraphics[width=65 mm]{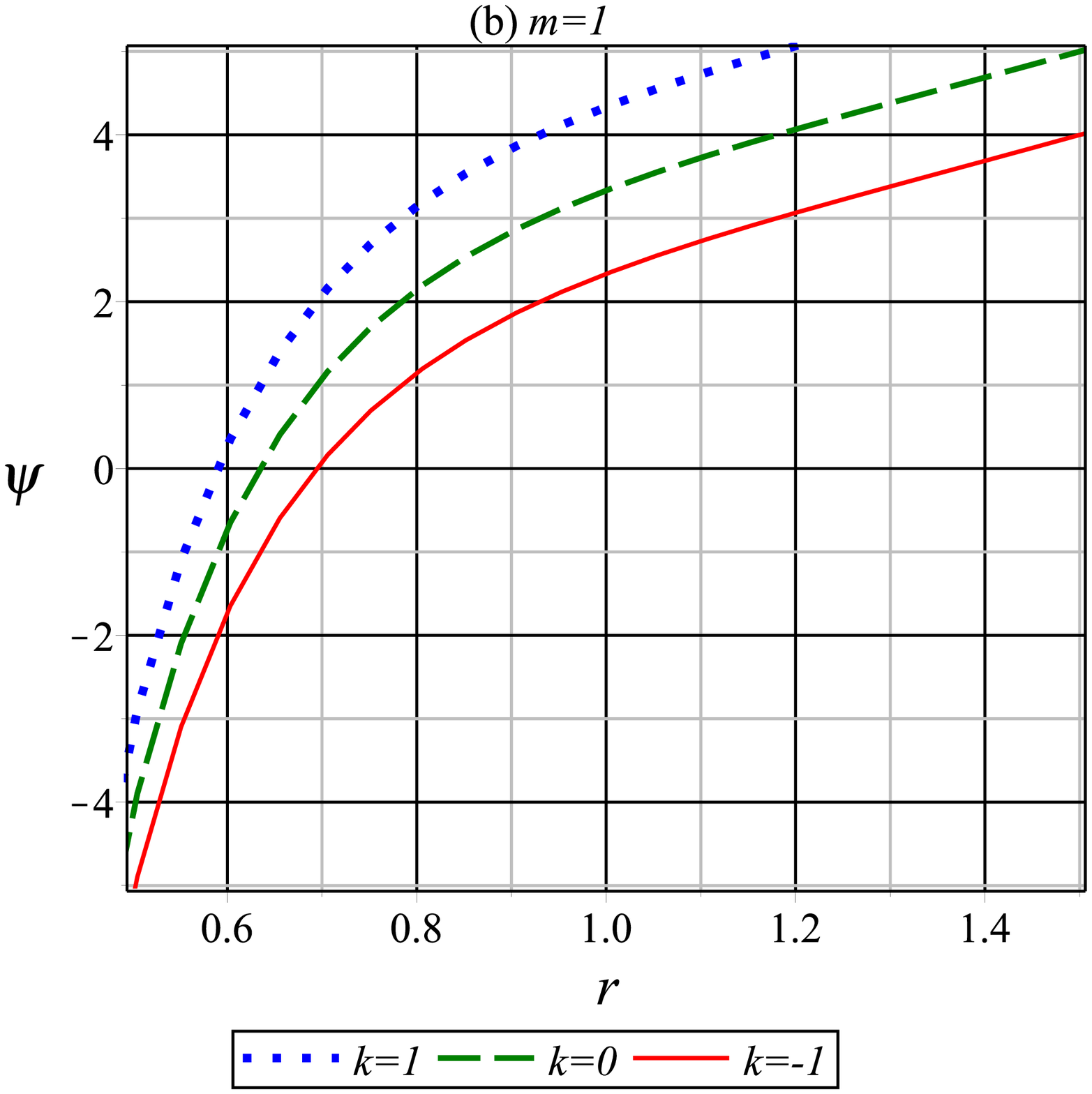}
&  &  &
\end{array}%
$%
\end{center}
\caption{Horizon structure for unit values of the model parameters.}
\label{fig1}
\end{figure}

Solving the nonzero components of the field equation (such as Eq. (\ref{FE11}%
)), yields
\begin{equation*}
\psi \left( r\right) =k-\frac{m_{0}}{r^{2}}+\frac{1}{g^{2}\left( \varepsilon
\right) r^{2}}\left[ \frac{r^{4}}{l^{2}}-2e^{2}\ln \left( \frac{r}{L}\right)
+m^{2}\left( \frac{c_{0}c_{1}}{3}r^{3}+c_{0}^{2}c_{2}r^{2}+2c_{0}^{3}c_{3}r%
\right) \right] ,
\end{equation*}%
where $m_{0}$ is the mass parameter, which is related to the black hole
mass, and $L$ is a constant with length dimension, introduced to obtain
dimensionless logarithmic function (which we can set to one without loss of
generality). Horizon structure of this solution shows that there is at least
one real positive root for $\psi (r)=0$ which is confirmed by the plots of
Fig. \ref{fig1}. In Fig. \ref{fig1} (a) we analyze the effect of mass term
in the massive gravity. Effects of other parameters like $c_{0}$, $e$ and $%
c_{1}$ are illustrated by plots of Fig. \ref{fig1} (b), (c) and (d),
respectively. In order to plot, we assume $c_{1}\approx c_{2}\approx c_{3}$,
for simplicity.

In the case of $r=r_{+}$, we have $\psi \left( r_{+}\right) =0$, and hence
\begin{equation*}
m_{0}=r_{+}^{2}\left( k+\frac{1}{g^{2}\left( \varepsilon \right) r_{+}^{2}}%
\left[ \frac{r_{+}^{4}}{l^{2}}-2e^{2}\ln (r_{+})+m^{2}\left( \frac{c_{0}c_{1}%
}{3}r_{+}^{3}+c_{0}^{2}c_{2}r_{+}^{2}+2c_{0}^{3}c_{3}r_{+}\right) \right]
\right)
\end{equation*}%
Therefore, we can write
\begin{eqnarray}
\psi \left( r\right) &=&\frac{k}{r^{2}}\left( r^{2}-r_{+}^{2}\right) +\frac{1%
}{g^{2}\left( \varepsilon \right) r^{2}}\left[ \frac{r^{4}-r_{+}^{4}}{l^{2}}%
-2e^{2}\ln (\frac{r}{r_{+}})\right]  \notag  \label{eq21} \\
&+&\frac{m^{2}}{g^{2}\left( \varepsilon \right) r^{2}}\left[ \left( \frac{%
c_{0}c_{1}}{3}\left( r^{3}-r_{+}^{3}\right) +c_{0}^{2}c_{2}\left(
r^{2}-r_{+}^{2}\right) +2c_{0}^{3}c_{3}\left( r-r_{+}\right) \right) \right]
.
\end{eqnarray}%
Now, we can analyze the thermodynamics of this black hole solution as its
flow with scale.

\section{Thermodynamics}

\label{sec3}

In this section, we discuss the scale dependence of the thermodynamics of
this Yang-Mills black hole solution given by (\ref{eq21}). This will be done
using the formalism of gravity's rainbow. So, first of all we will discuss
general formalism, and then discuss certain special models, with specific
rainbow functions $f\left(\varepsilon\right)$ and $g\left(\varepsilon\right)$%
.

\subsection{General Formalism}

\label{sec3-1} In this section, we will review the general formalism for
black hole thermodynamics in gravity's rainbow \cite{faiz1, faiz2}. Hawking
temperature for a black hole in an energy-dependent metric is given by
\begin{equation}
T_{H}=\frac{1}{4\pi }\frac{g\left( \varepsilon \right) }{f\left( \varepsilon
\right) }\psi ^{^{\prime }}\left( r\right) _{r=r_{+}}.  \label{TH}
\end{equation}%
Thus, we obtain the following temperature for the Yang-Mills black hole
\begin{equation}
T_{H}=\frac{1}{4\pi }\left[ \frac{2k}{r_{+}}\frac{g\left( \varepsilon
\right) }{f\left( \varepsilon \right) }+\frac{1}{f\left( \varepsilon \right)
g\left( \varepsilon \right) }\left[ \frac{4r_{+}}{l^{2}}-\frac{2e^{2}}{%
r_{+}^{3}}+m^{2}\left( c_{0}c_{1}+2\frac{c_{0}^{2}c_{2}}{r_{+}}+2\frac{%
c_{0}^{3}c_{3}}{r_{+}^{2}}\right) \right] \right] .  \label{Temp}
\end{equation}%
Also, the entropy per unit volume $\omega _{k}$ is given by
\begin{equation*}
S=\frac{r_{+}^{3}}{4g^{3}\left( \varepsilon \right) }.
\end{equation*}%
So, the mass term per unit volume $\omega _{k}$ can be obtained from the
first law of black hole thermodynamics
\begin{equation*}
T=\left( \frac{\partial M}{\partial S}\right) .
\end{equation*}%
Therefore, we can write
\begin{equation*}
M=\int TdS,
\end{equation*}%
Thus, for the Yang-Mills theory, we obtain
\begin{eqnarray}
M &=&\frac{3k}{16\pi f\left( \varepsilon \right) g^{2}\left( \varepsilon
\right) }r_{+}^{2}  \notag  \label{eq28} \\
&+&\frac{3}{16\pi f\left( \varepsilon \right) g^{4}\left( \varepsilon
\right) }\left[ \frac{r_{+}^{4}}{l^{2}}-2e^{2}\ln \left( r_{+}\right)
+m^{2}\left( \frac{c_{0}c_{1}}{3}%
r_{+}^{3}+c_{0}^{2}c_{2}r_{+}^{2}+2c_{0}^{3}c_{3}r_{+}\right) \right] .
\end{eqnarray}%
The heat capacity at constant volume can be calculated as
\begin{equation*}
C=T_{H}\left( \frac{\partial S}{\partial T_{H}}\right) _{V},
\end{equation*}%
which yields to the following expression
\begin{equation*}
C=\frac{3g^{3}\left( \varepsilon \right) r_{+}^{2}}{4}\frac{\left[ \frac{2k}{%
r_{+}}\frac{g\left( \varepsilon \right) }{f\left( \varepsilon \right) }+%
\frac{1}{f\left( \varepsilon \right) g\left( \varepsilon \right) }\left[
\frac{4r_{+}}{l^{2}}-\frac{2e^{2}}{r_{+}^{3}}+m^{2}\left( c_{0}c_{1}+2\frac{%
c_{0}^{2}c_{2}}{r_{+}}+2\frac{c_{0}^{3}c_{3}}{r_{+}^{2}}\right) \right] %
\right] }{\left[ -\frac{2k}{r_{+}^{2}}\frac{g\left( \varepsilon \right) }{%
f\left( \varepsilon \right) }+\frac{1}{f\left( \varepsilon \right) g\left(
\varepsilon \right) }\left[ \frac{4}{l^{2}}+\frac{6e^{2}}{r_{+}^{4}}%
+m^{2}\left( -2\frac{c_{0}^{2}c_{2}}{r_{+}^{2}}-4\frac{c_{0}^{3}c_{3}}{%
r_{+}^{3}}\right) \right] \right] }.
\end{equation*}%
It can be used to analyze the stability of specific models. If its sign be
positive the model is in the stable phase, and vice versa. It may be noted
that black hole's radius with $C=0$ is important, as at that stage the black
hole does not exchange any energy with the surroundings. If $C=0$, then we
obtain the following equation
\begin{equation*}  \label{eq33}
4r_{+}^{4}+m^{2}l^{2}c_{0}c_{1}r_{+}^{3}+2l^{2}\left(
m^{2}c_{0}^{2}c_{2}+kg^{2}\left( \varepsilon \right) \right)
r_{+}^{2}+2m^{2}l^{2}c_{0}^{3}c_{3}r_{+}-2l^{2}e^{2}=0.
\end{equation*}

We can observe that various thermodynamics quantities for the model, and its
thermodynamics stability, depend on the choice of the rainbow functions $%
g\left( \varepsilon \right) $ and $f\left( \varepsilon \right) $ \cite{AK,
ra12, ra14, ra18}. So, we will now analyze this model for the specific
choice of rainbow functions.

\subsection{Loop Quantum Gravity}

\label{sec3-2}

It has been observed that geometry of spacetime becomes energy-dependent in
loop quantum gravity, and the energy-dependent metric for such a spacetime
can be obtained using the following rainbow functions \cite{AC1, AC2}
\begin{eqnarray}  \label{eq34}
f\left(\varepsilon\right)=1 \\
g\left(\varepsilon\right)=\sqrt{1-\eta\varepsilon^{n}},
\end{eqnarray}
where $\eta$ is a dimensionless constant. It may be noted that these rainbow
functions are compatible with the results obtained from both loop quantum
gravity and non-commutative geometry \cite{AC2}. Without loss of generality,
we can set $\eta =1$.

\begin{figure}[h!]
\begin{center}
$%
\begin{array}{cccc}
\includegraphics[width=50 mm]{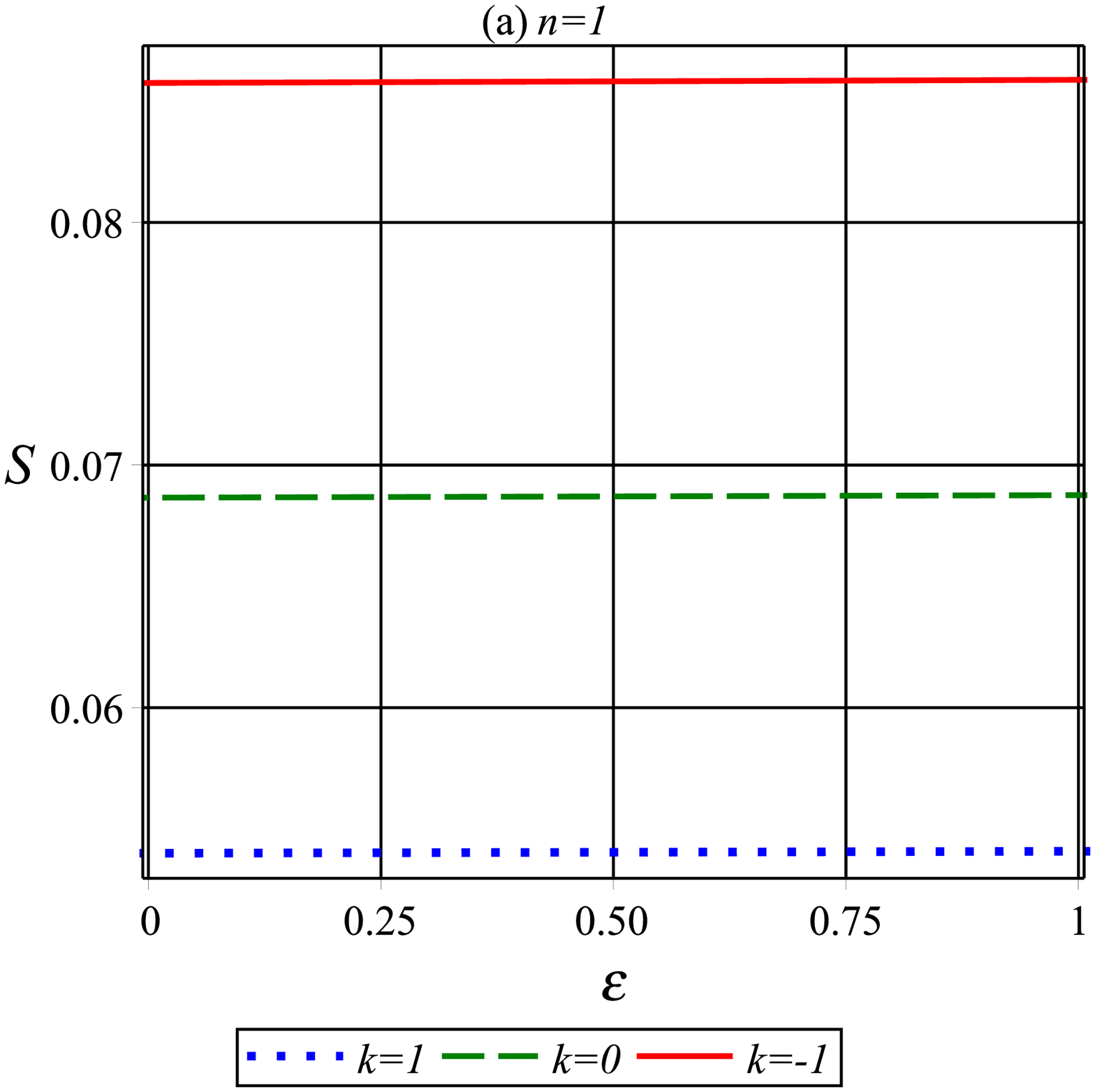}%
\includegraphics[width=50
mm]{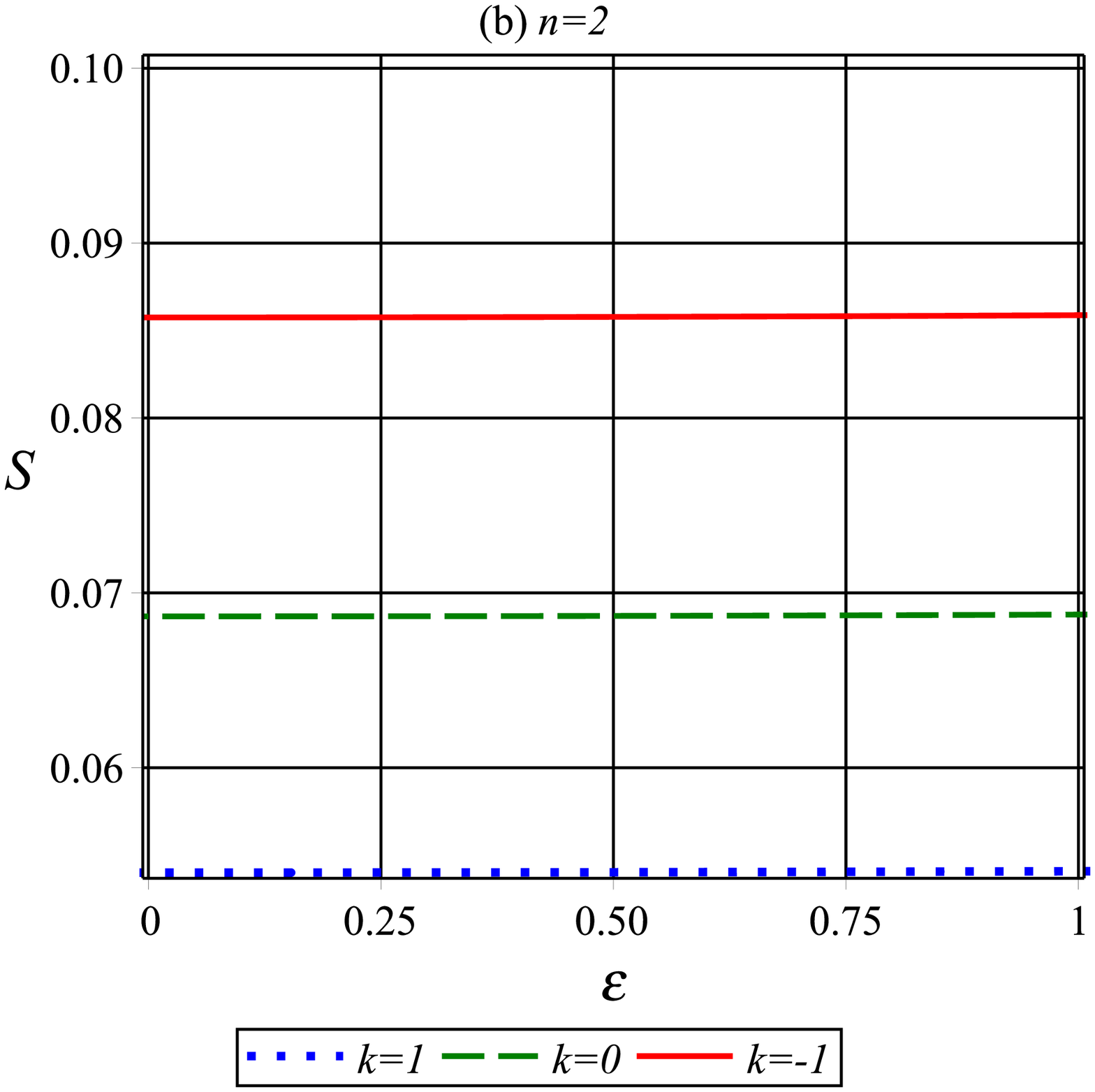}\includegraphics[width=50 mm]{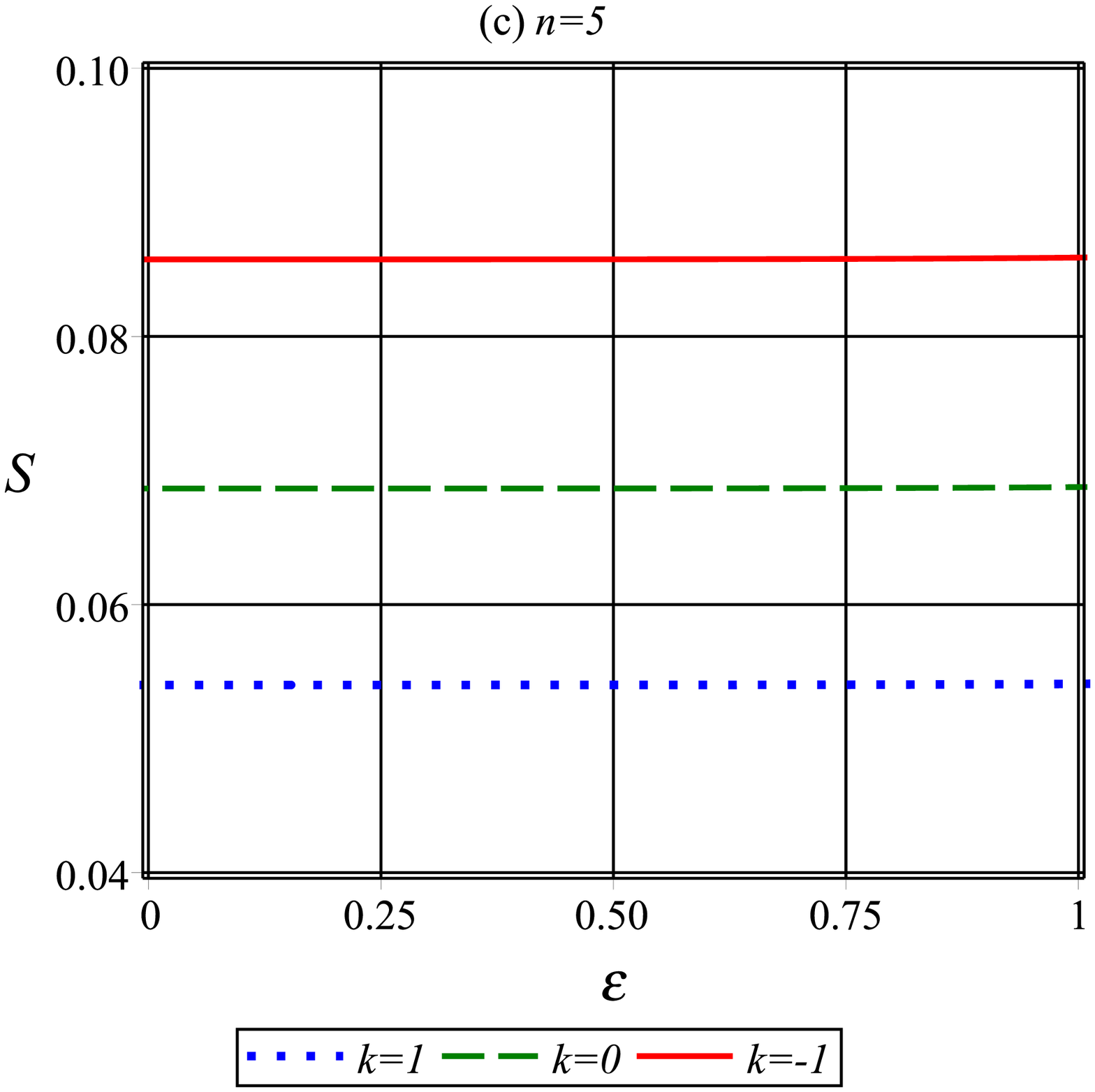} &  &  &
\end{array}%
$%
\end{center}
\caption{Entropy of the first model (equation (\protect\ref{eq36})) in terms
of $\protect\varepsilon$ for $\protect\eta=0.001$ and $r_{+}=0.5$.}
\label{figS-1}
\end{figure}

In this model, we can obtain entropy as
\begin{eqnarray}  \label{eq36}
S=\frac{r_{+}^{3}}{4\left(1-\eta\varepsilon^{n}\right)^{\frac{3}{2}}}.
\end{eqnarray}
Now according to Fig. \ref{fig1} (b), we can approximate $0.6\leq
r_{+}\leq0.7$, for unit values of the model parameters. Hence, we can
discuss the entropy for various values of $n$ as plotted in Fig. \ref{figS-1}%
. It is obvious that for $\varepsilon$ the entropy is approximately
constant. Generally, it is an increasing function of $\varepsilon$, but
local behavior depends on the value of $n$.\newline

\begin{figure}[h!]
\begin{center}
$%
\begin{array}{cccc}
\includegraphics[width=50 mm]{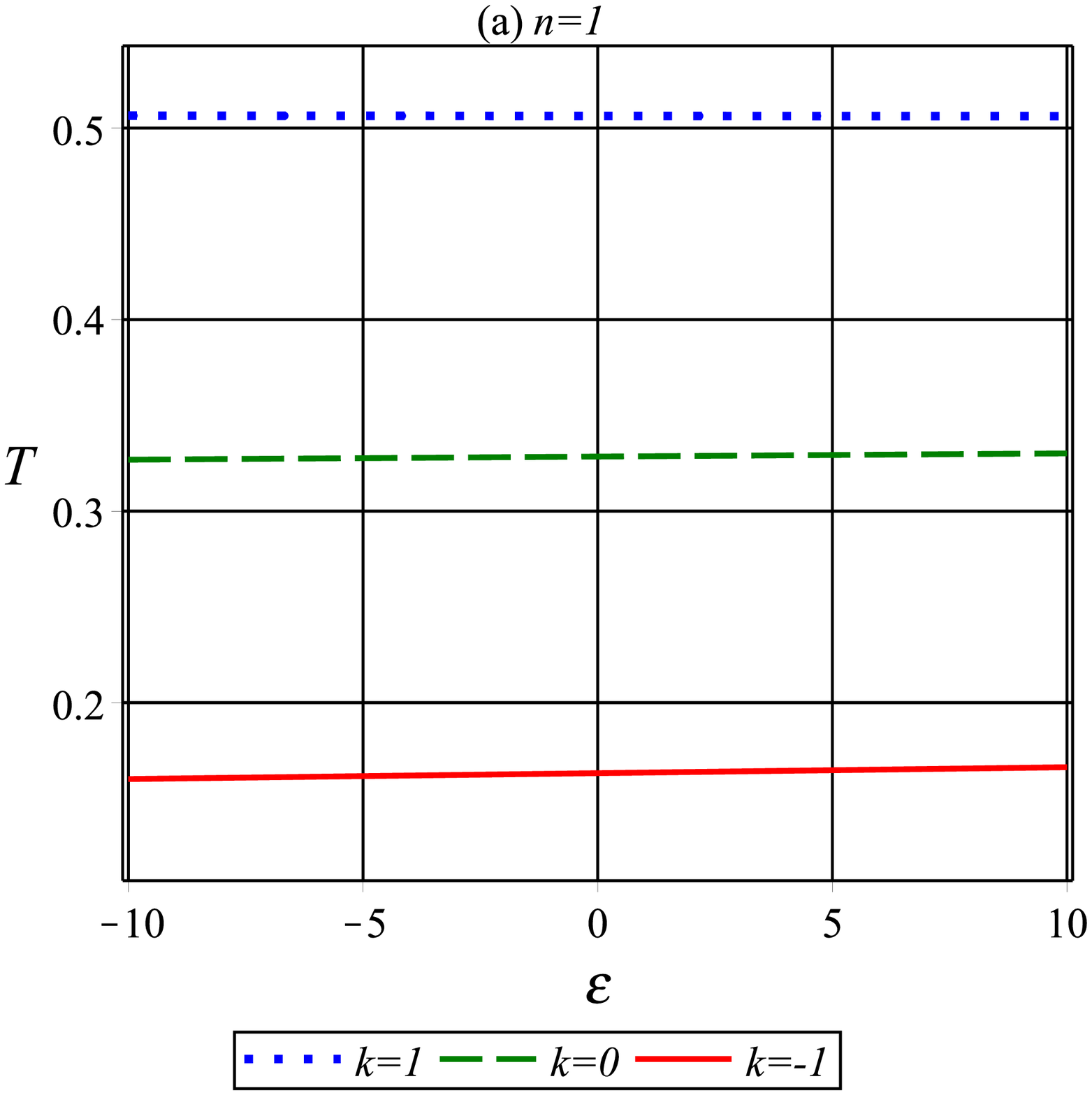}%
\includegraphics[width=50
mm]{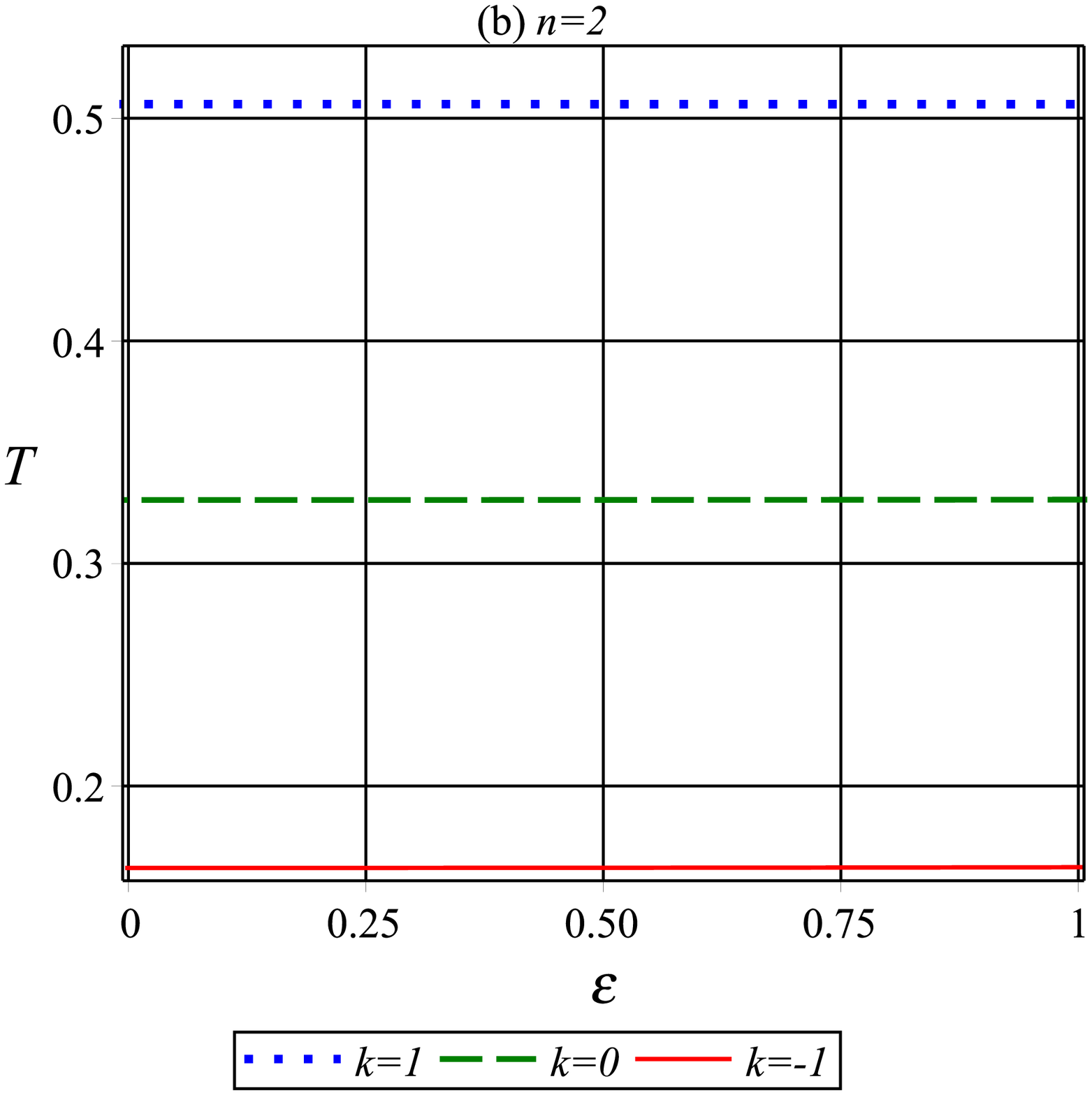}\includegraphics[width=50 mm]{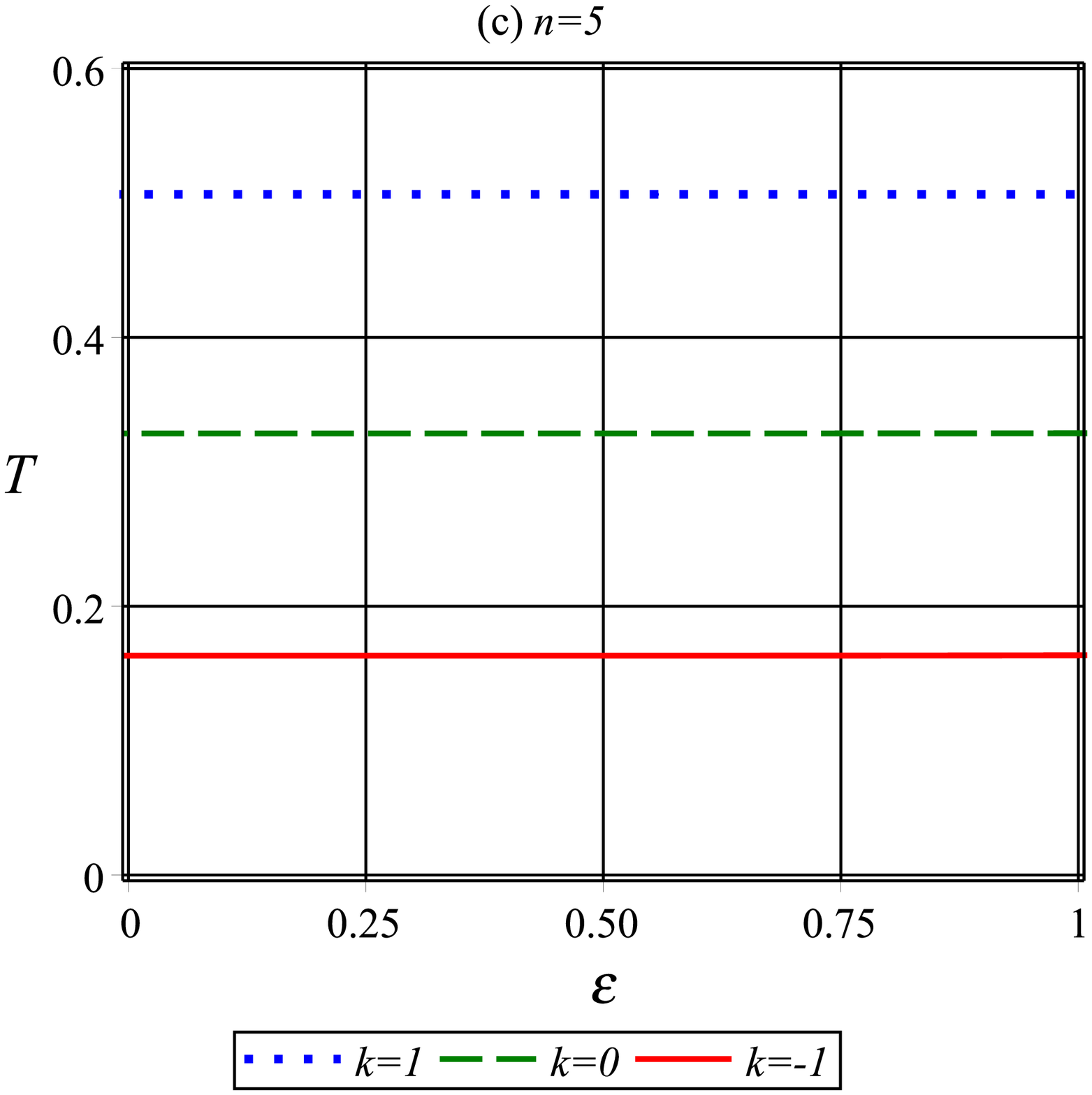} &  &  &
\end{array}%
$%
\end{center}
\caption{Temperature of the first model (equation (\protect\ref{eq37})) in
terms of $\protect\varepsilon$ for $\protect\eta=0.001$ and $r_{+}=0.5$
(unit value for other parameters).}
\label{figT-1}
\end{figure}

Temperature can be expressed as
\begin{eqnarray}  \label{eq37}
T_{H}=\frac{1}{4\pi}\left[\frac{2k}{r_{+}}\sqrt{1-\eta\varepsilon^{n}}+\frac{%
1}{\sqrt{1-\eta\varepsilon^{n}}}\left[\frac{4r_{+}}{l^{2}}-\frac{2e^{2}}{%
r_{+}^{3}}+m^{2}\left(c_{0}c_{1}+2\frac{c_{0}^{2}c_{2}}{r_{+}}+2\frac{%
c_{0}^{3}c_{3}}{r_{+}^{2}}\right)\right]\right]
\end{eqnarray}
Using positive temperature, we can find a lower bound for the graviton mass.
So, for small $m$ the temperature is negative, which is not physical. Hence,
the value of $m$ is bounded ($m\geq0.9$ for our selected values of model
parameters). As illustrated by the plots of Fig. \ref{figT-1}, temperature
is constant for small $\varepsilon$.\newline
Now for $n=1$ Fig. \ref{figT-1} (a) we find $T$ is an increasing function of
$\varepsilon$, for $k=-1$ and $k=0$, and a decreasing function of $%
\varepsilon$, for $k=1$ (with infinitesimal variation). So, the behavior of $%
T$ for $k=1$ is also different from $k=-1$ and $k=0$, for $n=1$ as
illustrated by Fig. \ref{figT-1} (b). We can also analyze the behavior of
the entropy for $\varepsilon$. \newline

The black hole mass for this model is given by
\begin{eqnarray}  \label{eq38}
M&=&\frac{3k r_{+}^{2}}{16\pi\left(1-\eta\varepsilon^{n}\right)}  \notag \\
&+&\frac{3}{16\pi\left(1-\eta\varepsilon^{n}\right)^{2}}\left[\frac{r_{+}^{4}%
}{l^{2}}-2e^{2}\ln\left(r_{+}\right) +m^{2}\left(\frac{c_{0}c_{1}}{3}%
r_{+}^{3}+c_{0}^{2}c_{2}r_{+}^{2}+2c_{0}^{3}c_{3}r_{+}\right)\right].
\end{eqnarray}

\begin{figure}[h!]
\begin{center}
$%
\begin{array}{cccc}
\includegraphics[width=50 mm]{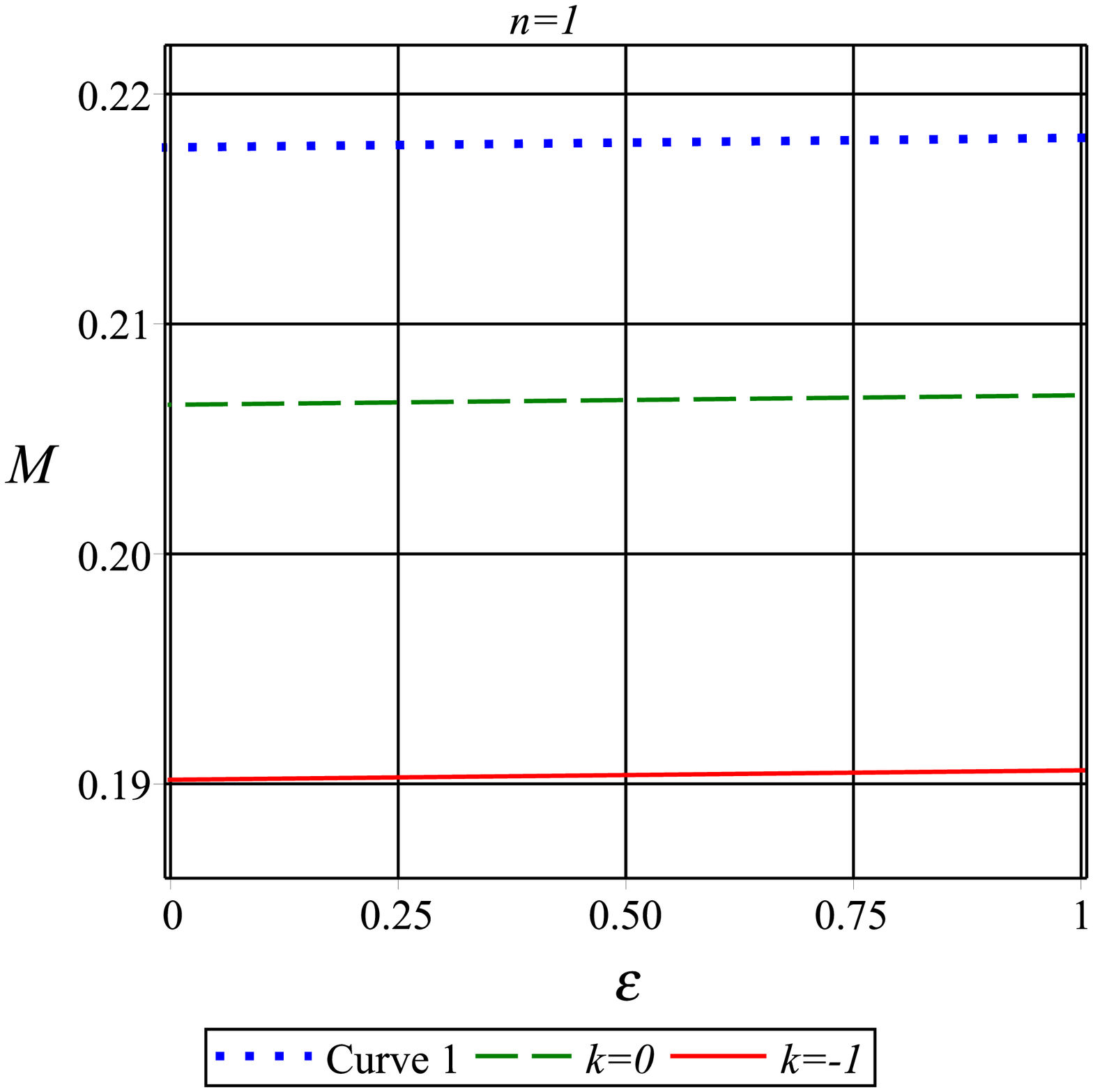}%
\includegraphics[width=50
mm]{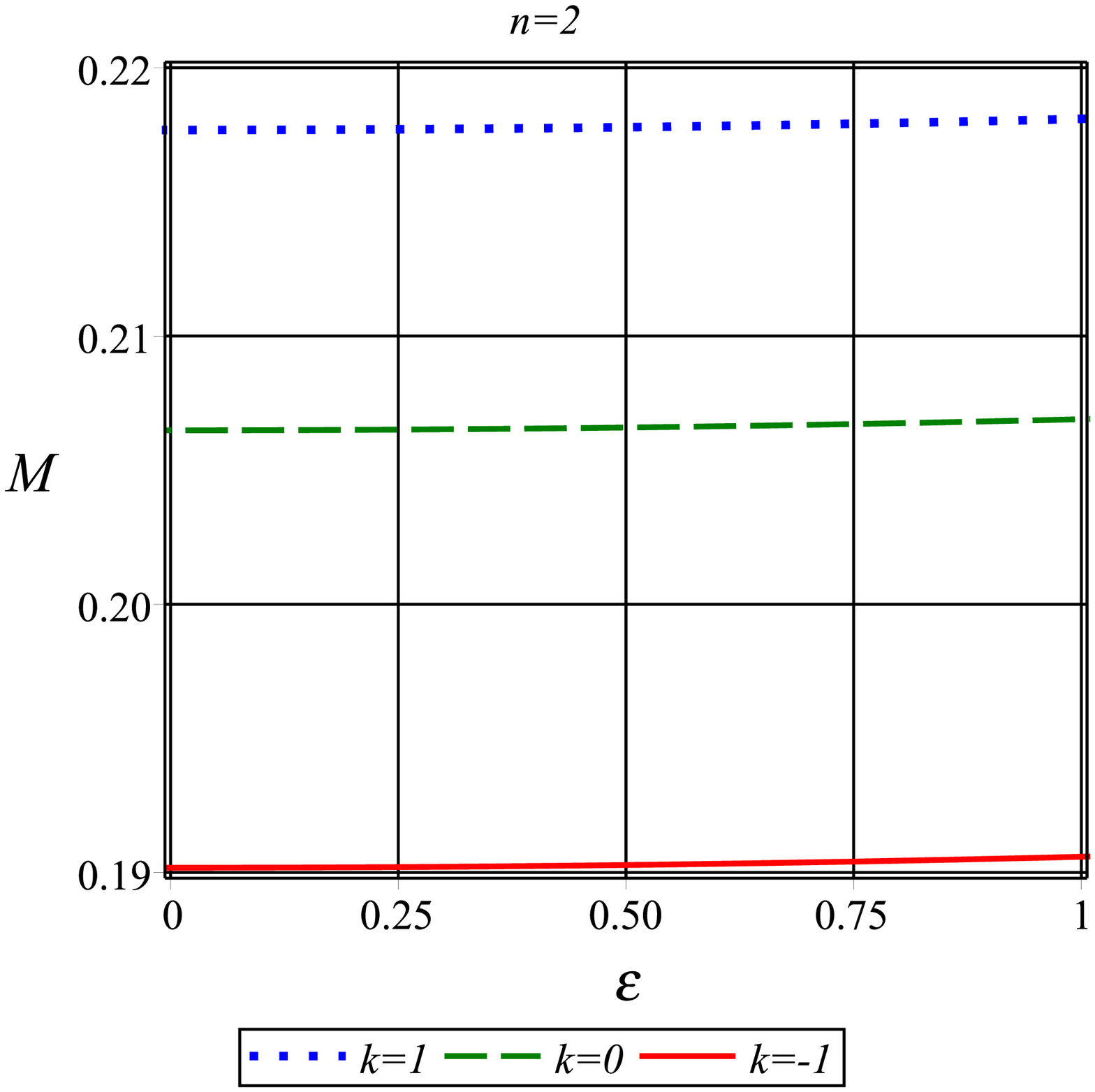}\includegraphics[width=50 mm]{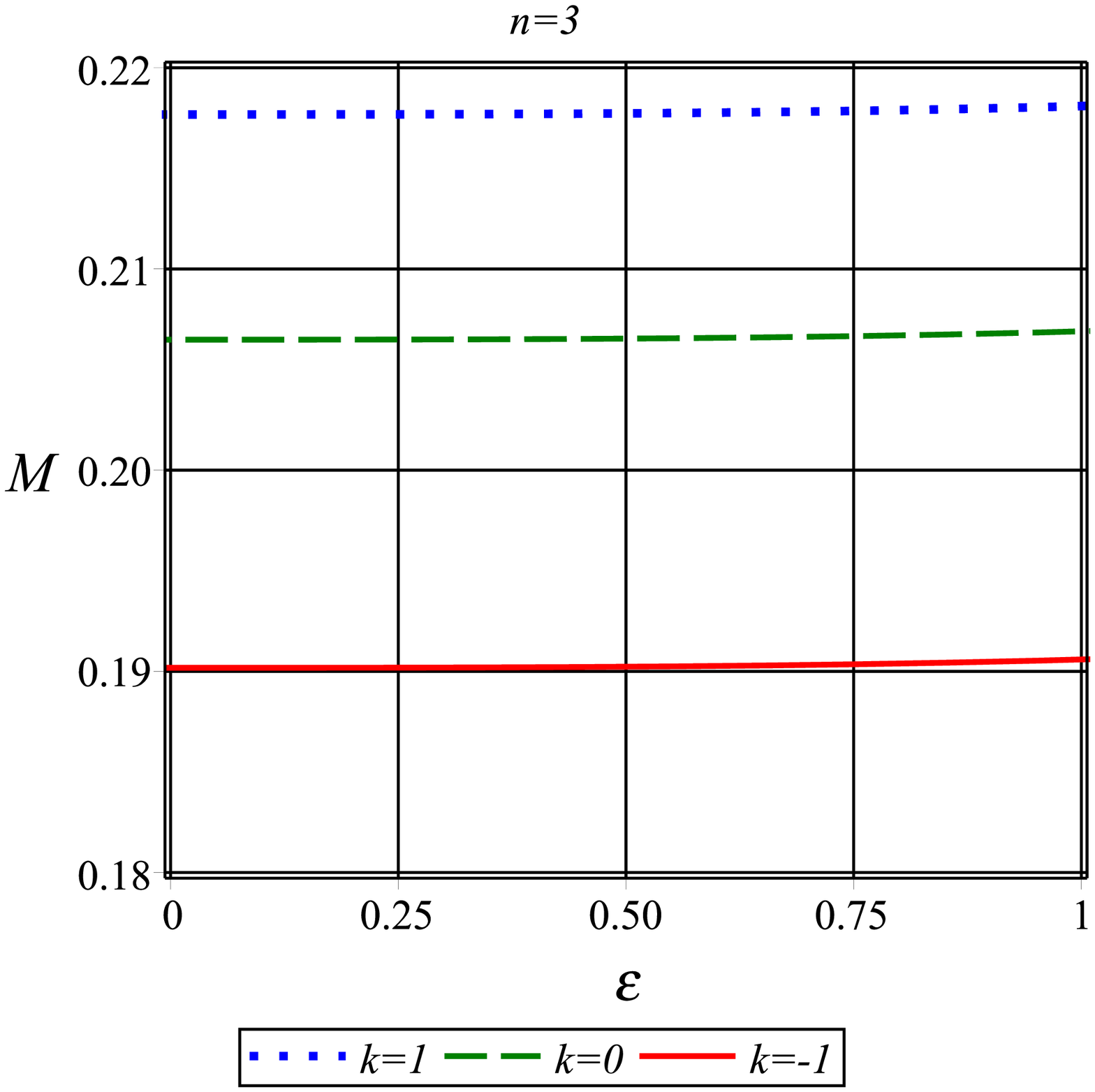} &  &  &  \\
\includegraphics[width=50 mm]{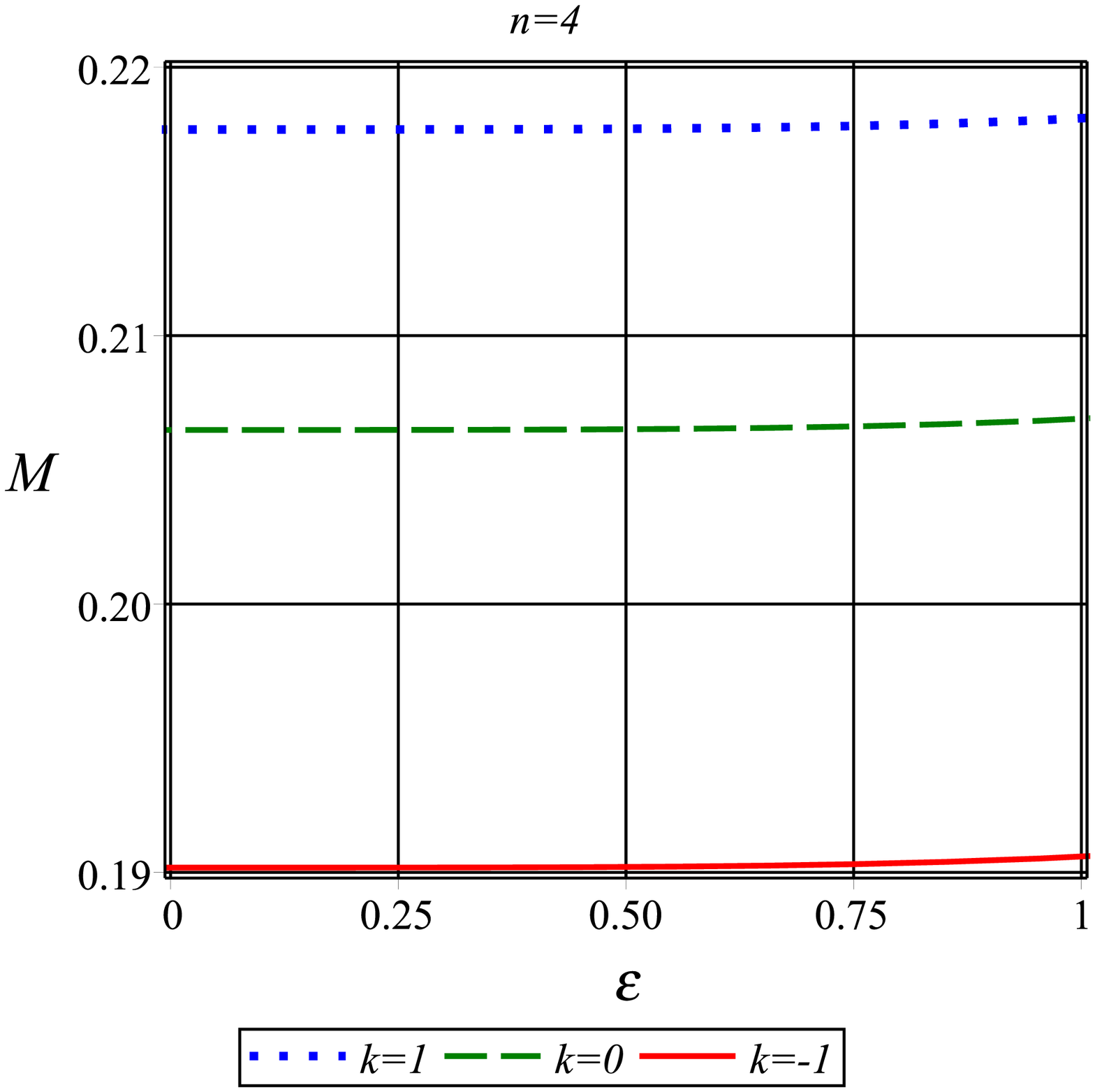}%
\includegraphics[width=50
mm]{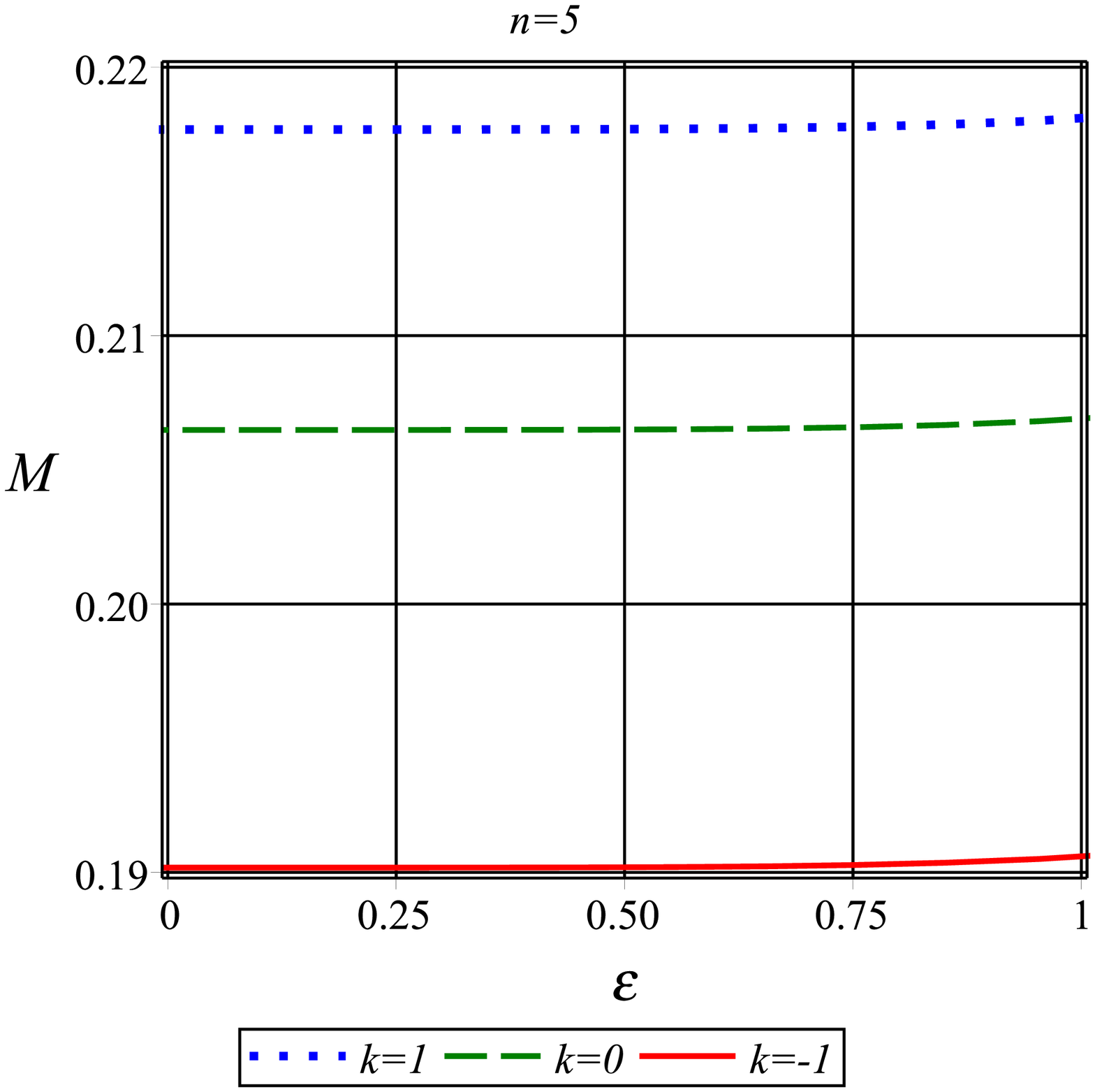}\includegraphics[width=50 mm]{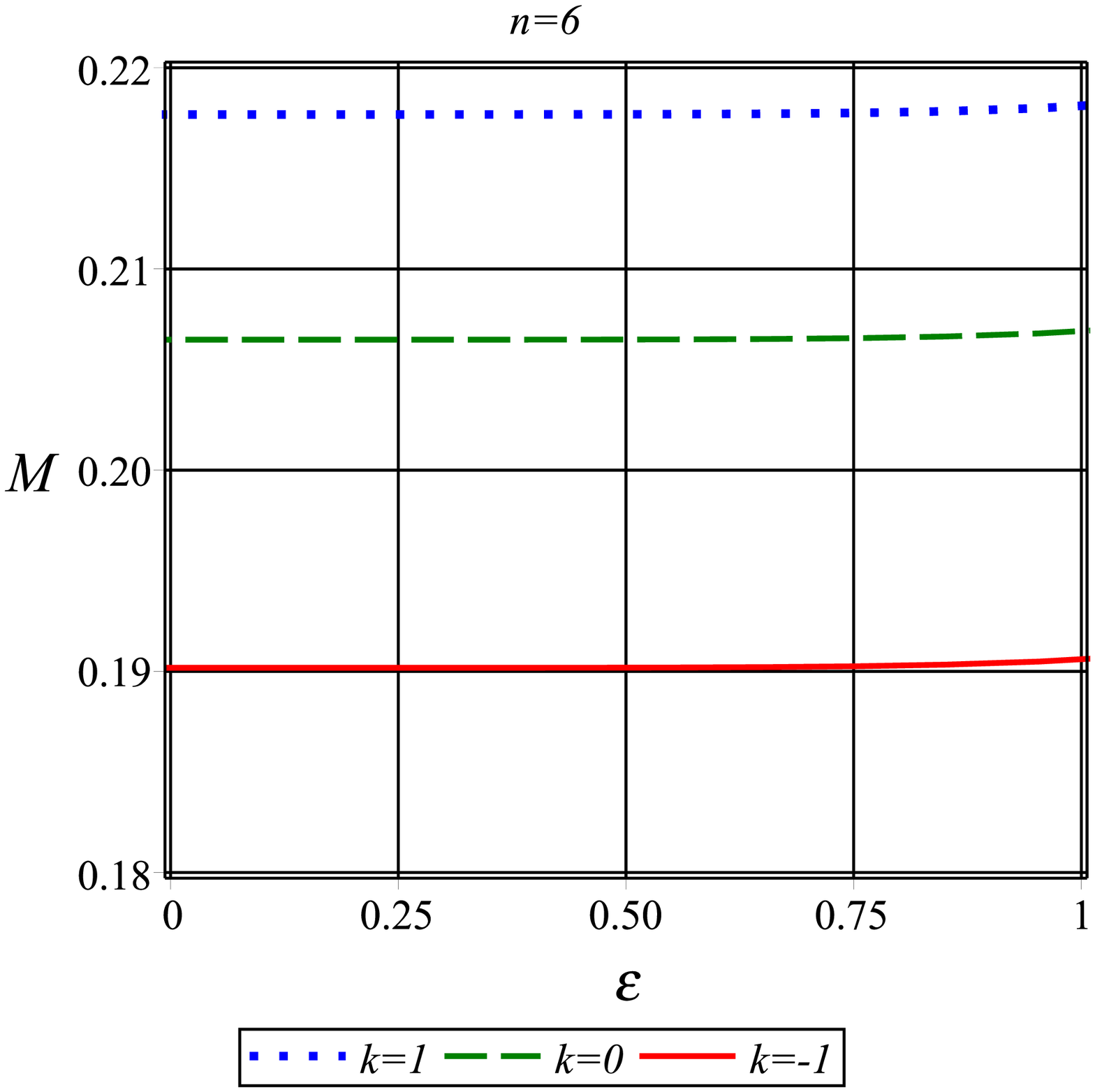} &  &  &
\end{array}%
$%
\end{center}
\caption{The first model mass (equation (\protect\ref{eq38})) in terms of $%
\protect\varepsilon$ for $\protect\eta=0.001$, and unit value for other
parameters.}
\label{figM-1}
\end{figure}

Typical values of the black hole mass given are given in Fig. \ref{figM-1},
for several values of $n$. For $n=1$ (linear dependence to energy), which is
consistent with results obtained from loop quantum gravity, we find that the
black hole mass may vanish for the negative $\varepsilon$ (see Fig. \ref%
{figM-1} (a)). For even values of $n$, we observe the periodic temperature
with a divergence point. So, there are some specific energies where the
black hole temperature diverges. It may be a sign of some instability or
phase transition, which can be verified from the specific heat of this
system.

Specific heat for this Yang-Mills black hole can now be expressed as
\begin{eqnarray}  \label{eq39}
C=\frac{3\left(1-\eta\varepsilon^{n}\right)^{\frac{3}{2}}r_{+}^{2}}{4}\frac{%
\left[\frac{2k}{r_{+}}\sqrt{1-\eta\varepsilon^{n}} +\frac{1}{\sqrt{%
1-\eta\varepsilon^{n}}}\left[\frac{4r_{+}}{l^{2}}-\frac{2e^{2}}{r_{+}^{3}}%
+m^{2}\left(c_{0}c_{1}+2\frac{c_{0}^{2}c_{2}}{r_{+}} +2\frac{c_{0}^{3}c_{3}}{%
r_{+}^{2}}\right)\right]\right]}{\left[-\frac{2k}{r_{+}^{2}} \sqrt{%
1-\eta\varepsilon^{n}}+\frac{1}{\sqrt{1-\eta\varepsilon^{n}}}\left[\frac{4}{%
l^{2}} +\frac{6e^{2}}{r_{+}^{4}}+m^{2}\left(-2\frac{c_{0}^{2}c_{2}}{r_{+}^{2}%
}-4\frac{c_{0}^{3}c_{3}}{r_{+}^{3}}\right)\right]\right]}.
\end{eqnarray}

\begin{figure}[h!]
\begin{center}
$%
\begin{array}{cccc}
\includegraphics[width=50 mm]{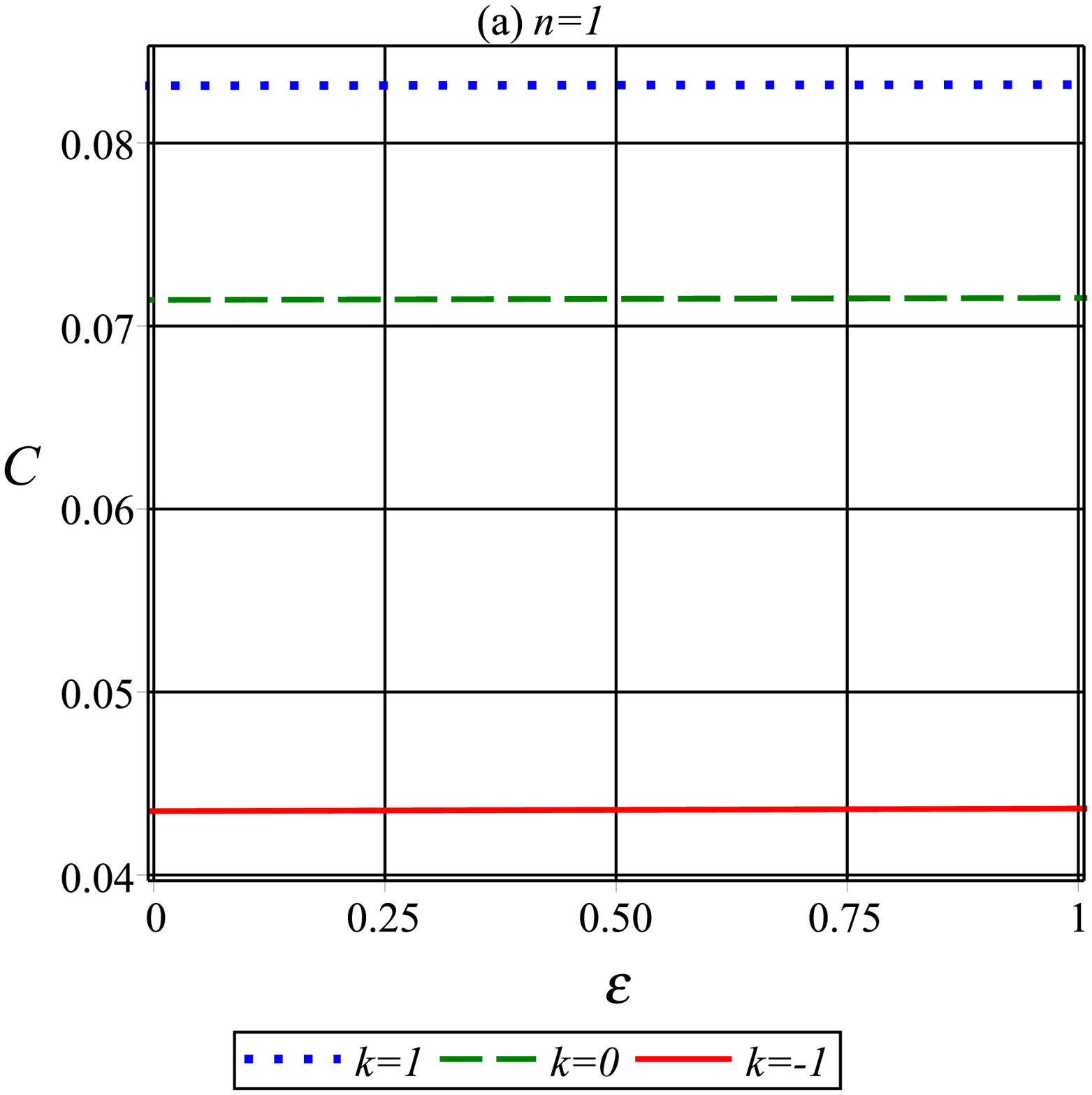}%
\includegraphics[width=50
mm]{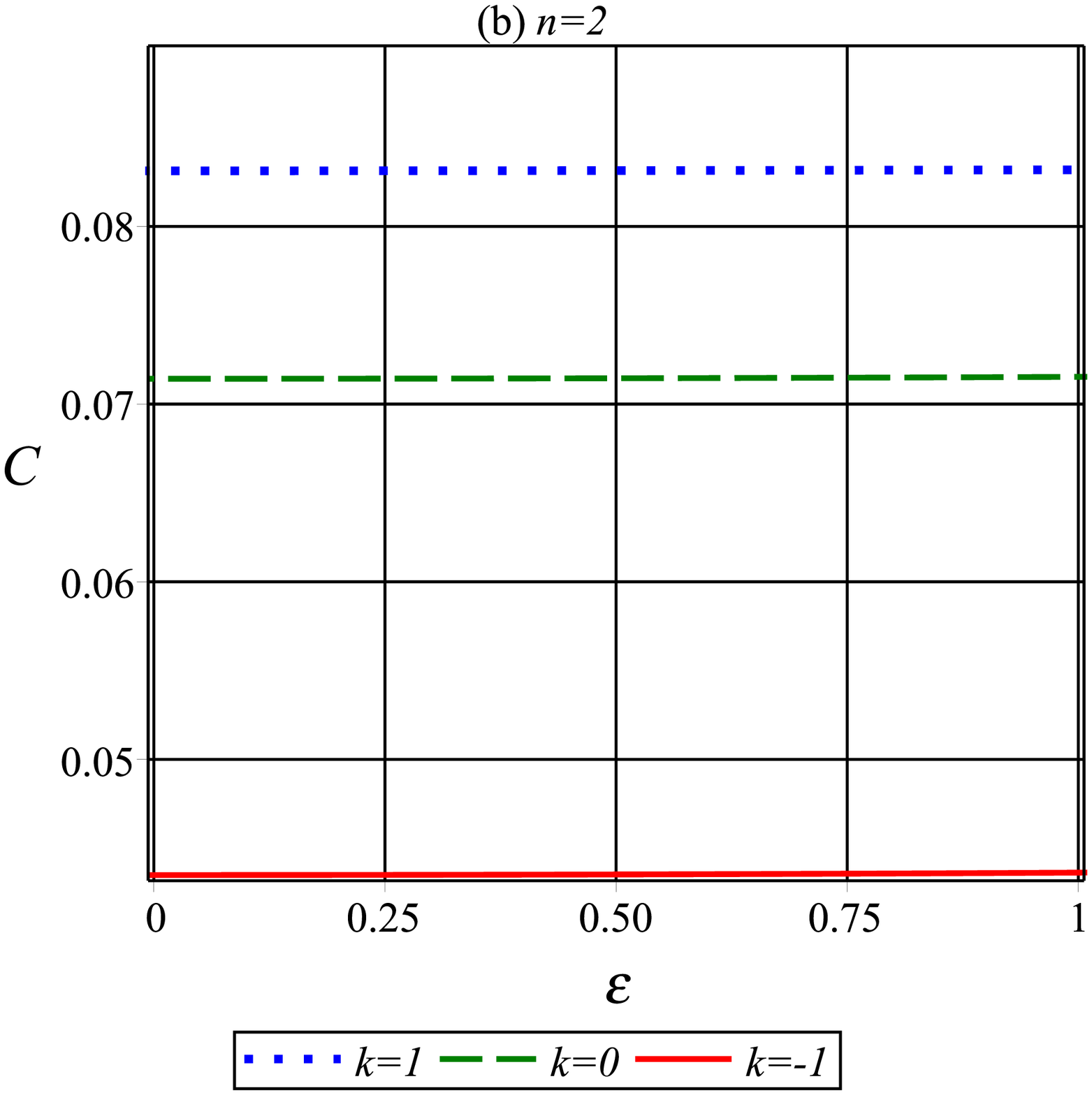}\includegraphics[width=50 mm]{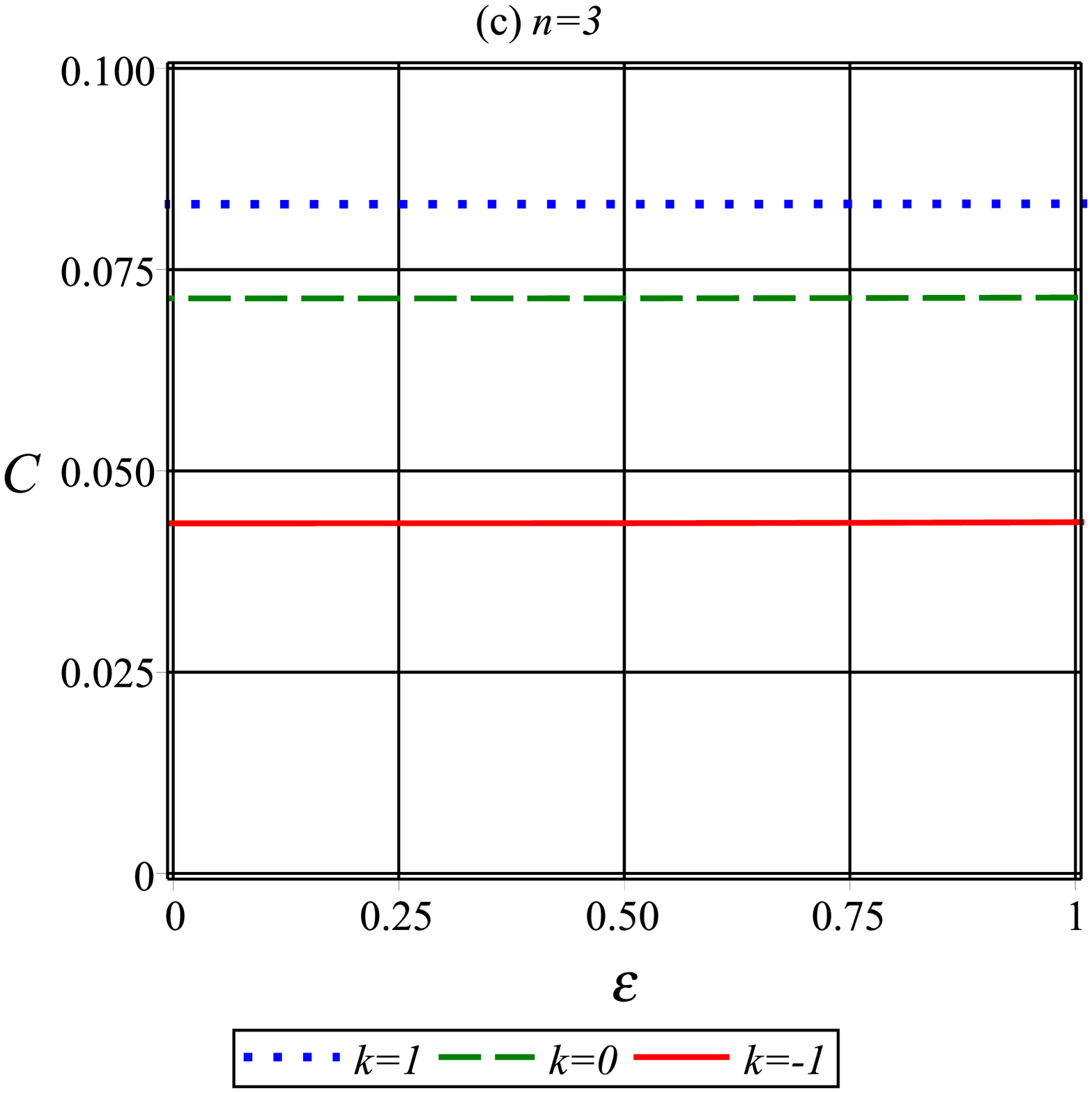} &  &  &  \\
\includegraphics[width=50 mm]{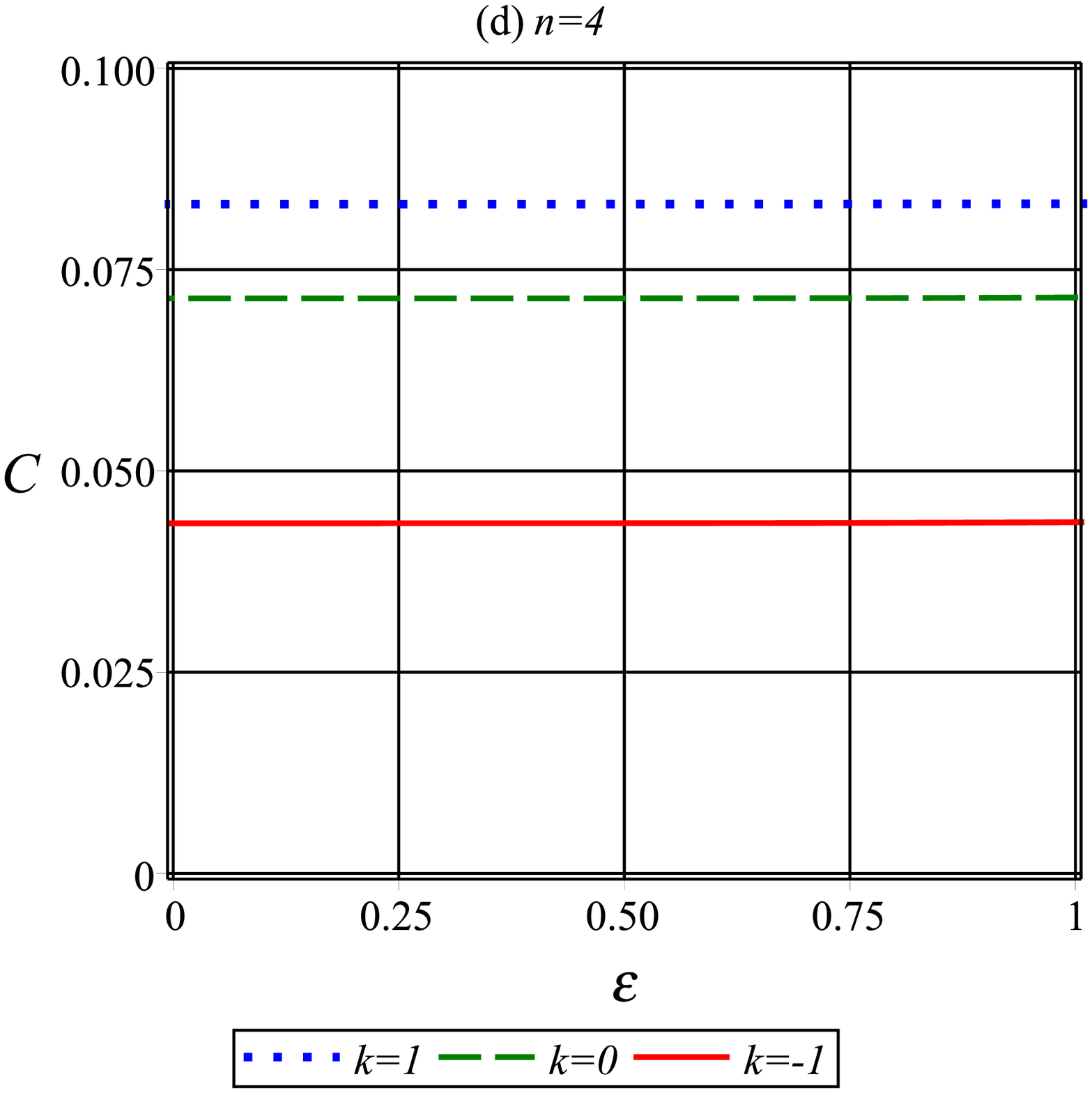}%
\includegraphics[width=50
mm]{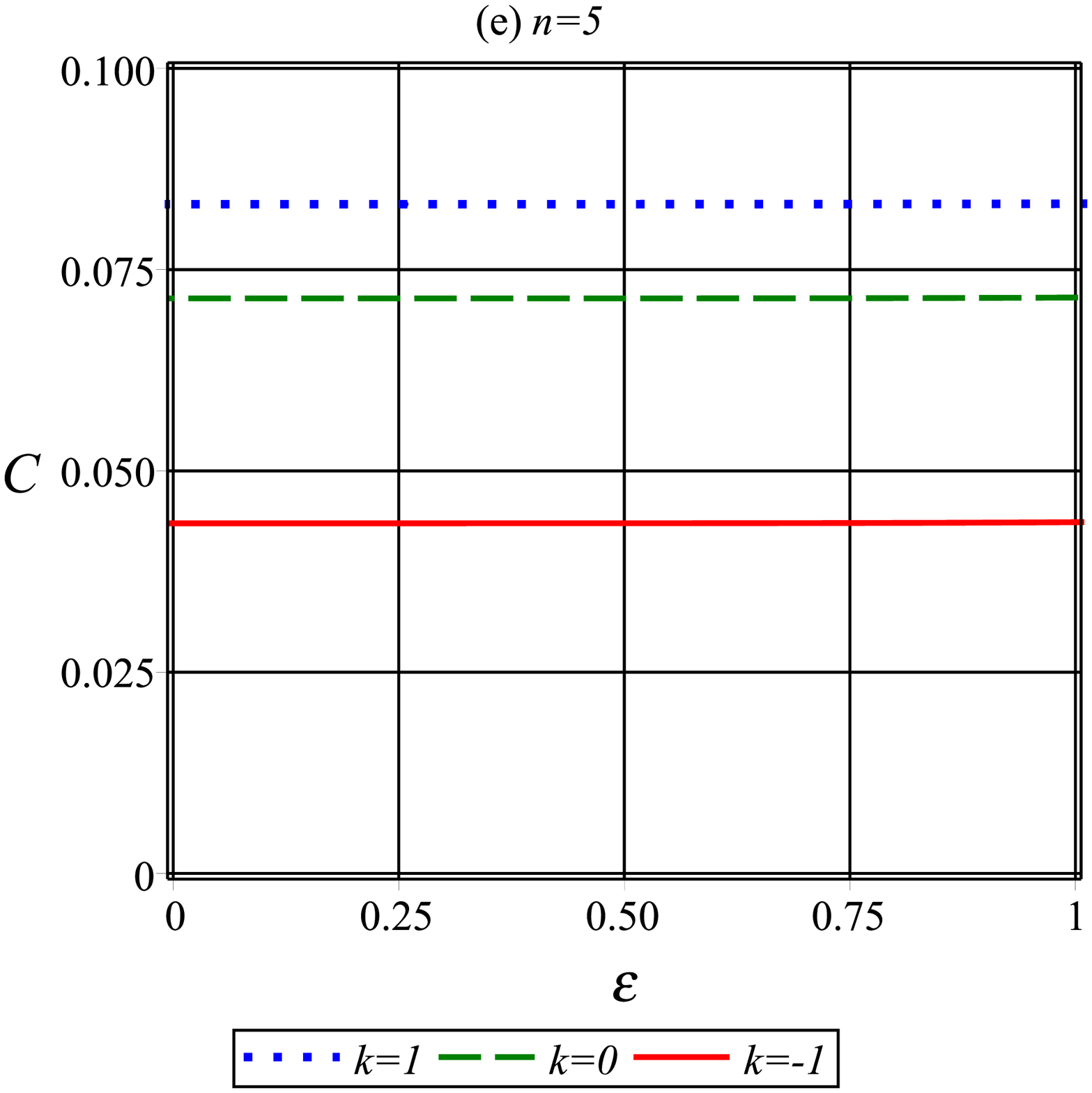}\includegraphics[width=50 mm]{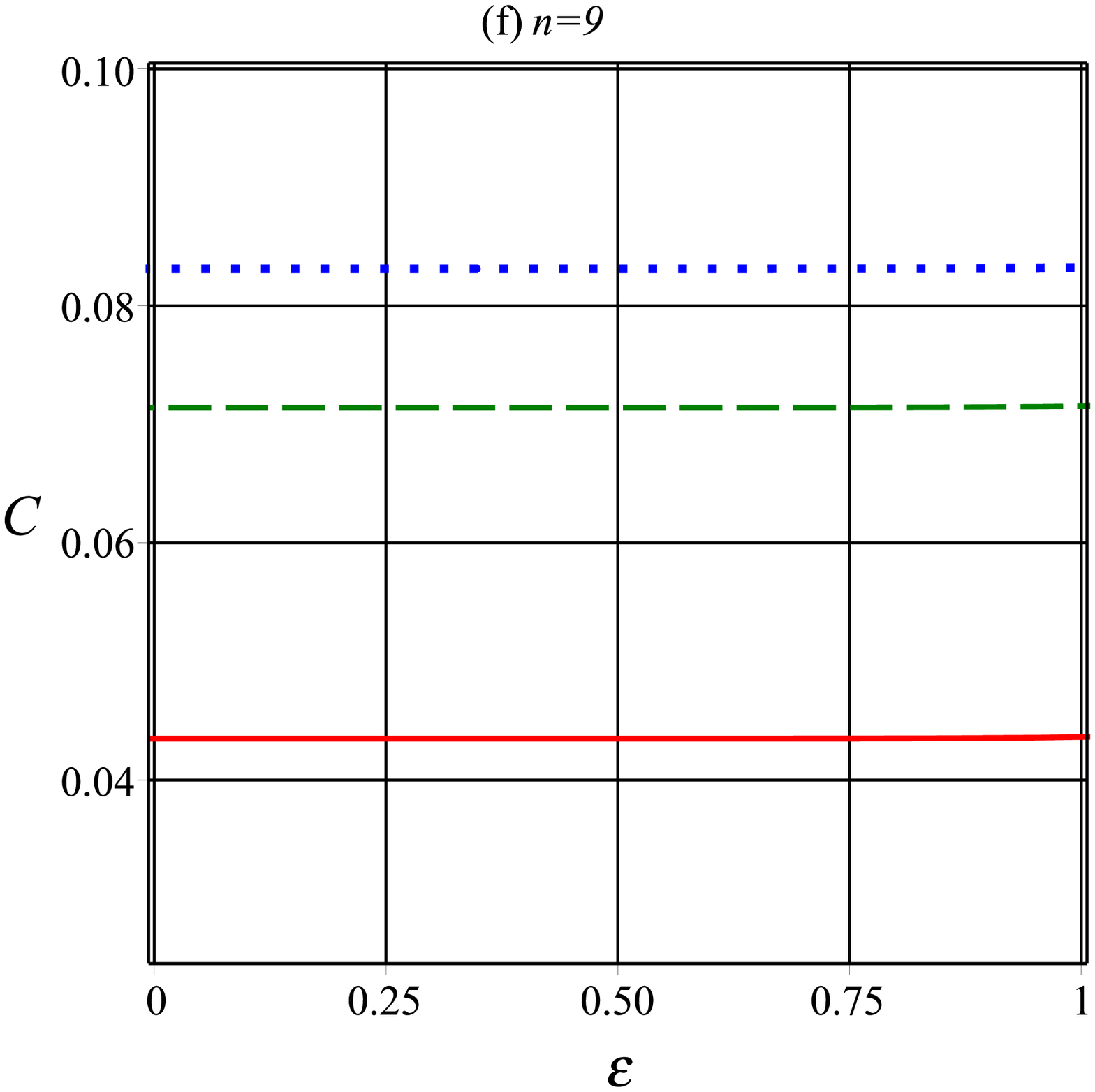} &  &  &
\end{array}%
$%
\end{center}
\caption{Specific heat of the first model (equation (\protect\ref{eq39})) in
terms of $\protect\varepsilon$ for $\protect\eta=0.001$, and unit value for
other parameters.}
\label{figC-1}
\end{figure}

As expected (from temperature as analyzed in Fig. \ref{figT-1}), specific
heat is negative for the small $m$. However, for the larger $m$, there are
some regions that are stable and others that are unstable. These are
illustrated in plots Fig. \ref{figC-1}. We found that in the cases of $m<1$,
the model is completely unstable as the specific heat is negative. However,
for the larger $m$, like $m\geq1$ (or $m\geq e$), the model is completely or
partly stable, and this depends on the value of $n$. Figs. \ref{figC-1} (a),
(b) and (d), which are corresponding to $n=1$, $n=2$ and $n=4$, show that
larger values of $m$ produce completely stable models. However, other odd
values of $n$ like $n=3$, $n=5$ and $n=9$ produce model with both stable and
unstable regions (depend on the value of $\varepsilon$). The specific heat
is negative for some negative value of $\varepsilon$ (see Figs. \ref{figC-1}
(c), (e) and (f)). As before, for the infinitesimal $\varepsilon$, the
specific heat is approximately constant. Figs. \ref{figC-1} (e) and (f),
indicating a phase transition for $k=1$. It is important to note that in the
range $0\leq\varepsilon\leq1$ the model is completely stable. \newline

Now, from Eq. (\ref{eq33}), we observe that for stability the following
condition is necessary
\begin{equation}  \label{eq40}
m>e.
\end{equation}
In fact, Eq. (\ref{eq33}) reduced to the following expression
\begin{eqnarray}  \label{eq333}
4r_{+}^{4}+m^{2}l^{2}c_{0}c_{1}r_{+}^{3}+2l^{2}\left(m^{2}c_{0}^{2}c_{2}+k-%
\eta
k\varepsilon^{n}%
\right)r_{+}^{2}+2m^{2}l^{2}c_{0}^{3}c_{3}r_{+}-2l^{2}e^{2}=0.
\end{eqnarray}
An exact bound on the energy for a stable model can now be written as
\begin{eqnarray}  \label{eq3333}
\varepsilon^{n}\geq\frac{%
2l^{2}e^{2}-4r_{+}^{4}-m^{2}l^{2}c_{0}c_{1}r_{+}^{3}-2l^{2}%
\left(m^{2}c_{0}^{2}c_{2}+k\right)r_{+}^{2}-2m^{2}l^{2}c_{0}^{3}c_{3}r_{+}} {%
-2kl^{2}\eta r_{+}^{2}}.
\end{eqnarray}
Hence, we find that for $E\ll E_{p}$, the first model is stable.

\subsection{Gamma-Ray Bursters}

\label{sec3-3} It is also possible to obtain an energy-dependent metric by
using rainbow function obtained from hard spectra of gamma-ray bursts at
cosmological distances \cite{AC3}
\begin{eqnarray}  \label{eq400}
f\left(\varepsilon\right)=\frac{e^{\xi\varepsilon}-1}{\xi\varepsilon}, \\
g\left(\varepsilon\right)=1,
\end{eqnarray}
where $\xi$ is a dimensionless parameters of the order of unity. In this
case, we have
\begin{eqnarray}  \label{eq42}
S=\frac{r_{+}^{3}}{4}.
\end{eqnarray}
We see that the entropy only depends on the horizon radius, and
\begin{eqnarray}  \label{eq43}
T_{H}=\frac{1}{4\pi}\left[\frac{2k}{r_{+}}\frac{\xi\varepsilon}{%
e^{\xi\varepsilon}-1} +\frac{\xi\varepsilon}{e^{\xi\varepsilon}-1}\left[%
\frac{4r_{+}}{l^{2}}-\frac{2e^{2}}{r_{+}^{3}} +m^{2}\left(c_{0}c_{1}+2\frac{
c_{0}^{2}c_{2}}{r_{+}}+2\frac{c_{0}^{3}c_{3}}{r_{+}^{2}}\right)\right]\right]%
.
\end{eqnarray}

\begin{figure}[h!]
\begin{center}
$%
\begin{array}{cccc}
\includegraphics[width=50 mm]{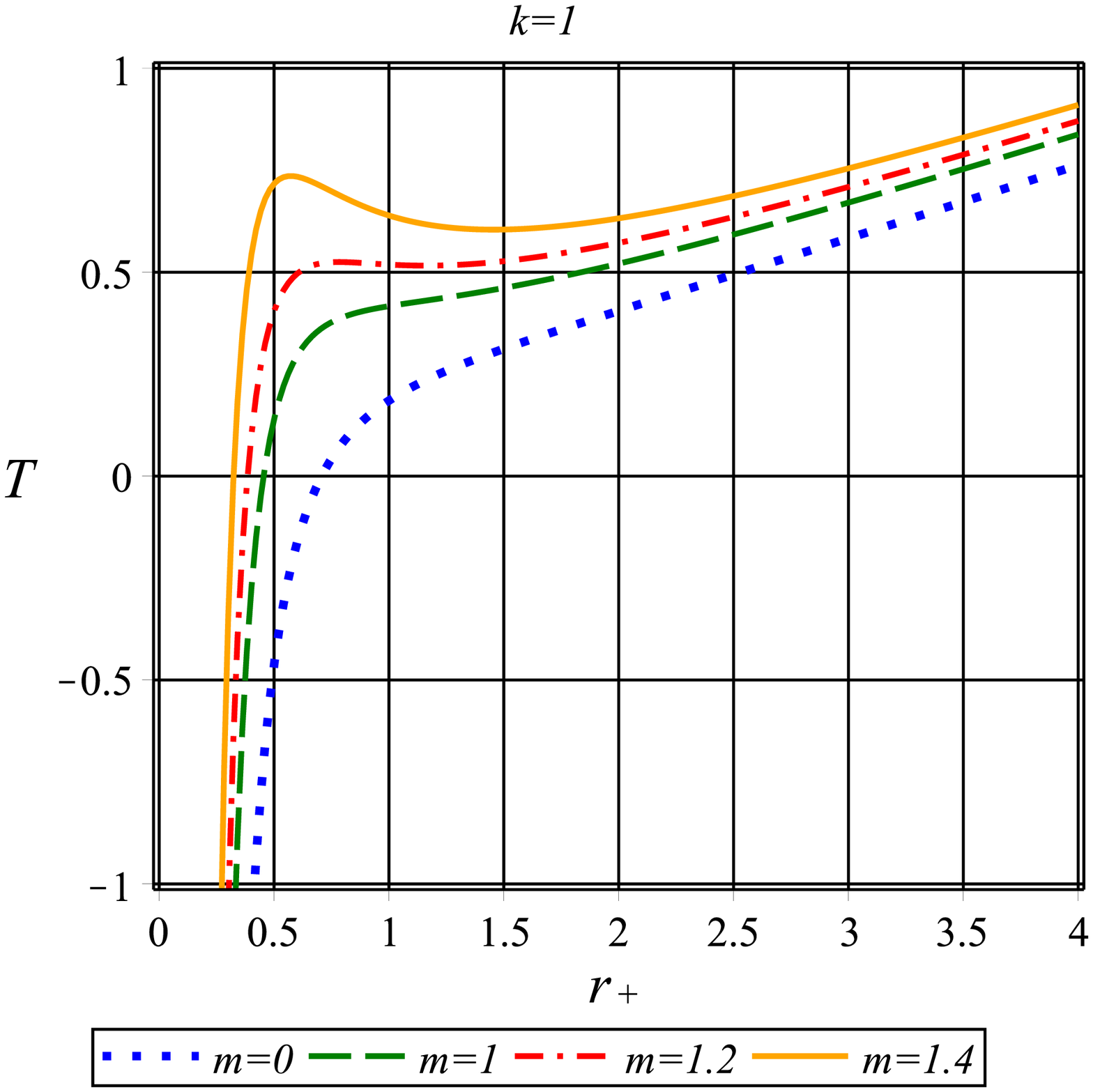}%
\includegraphics[width=50
mm]{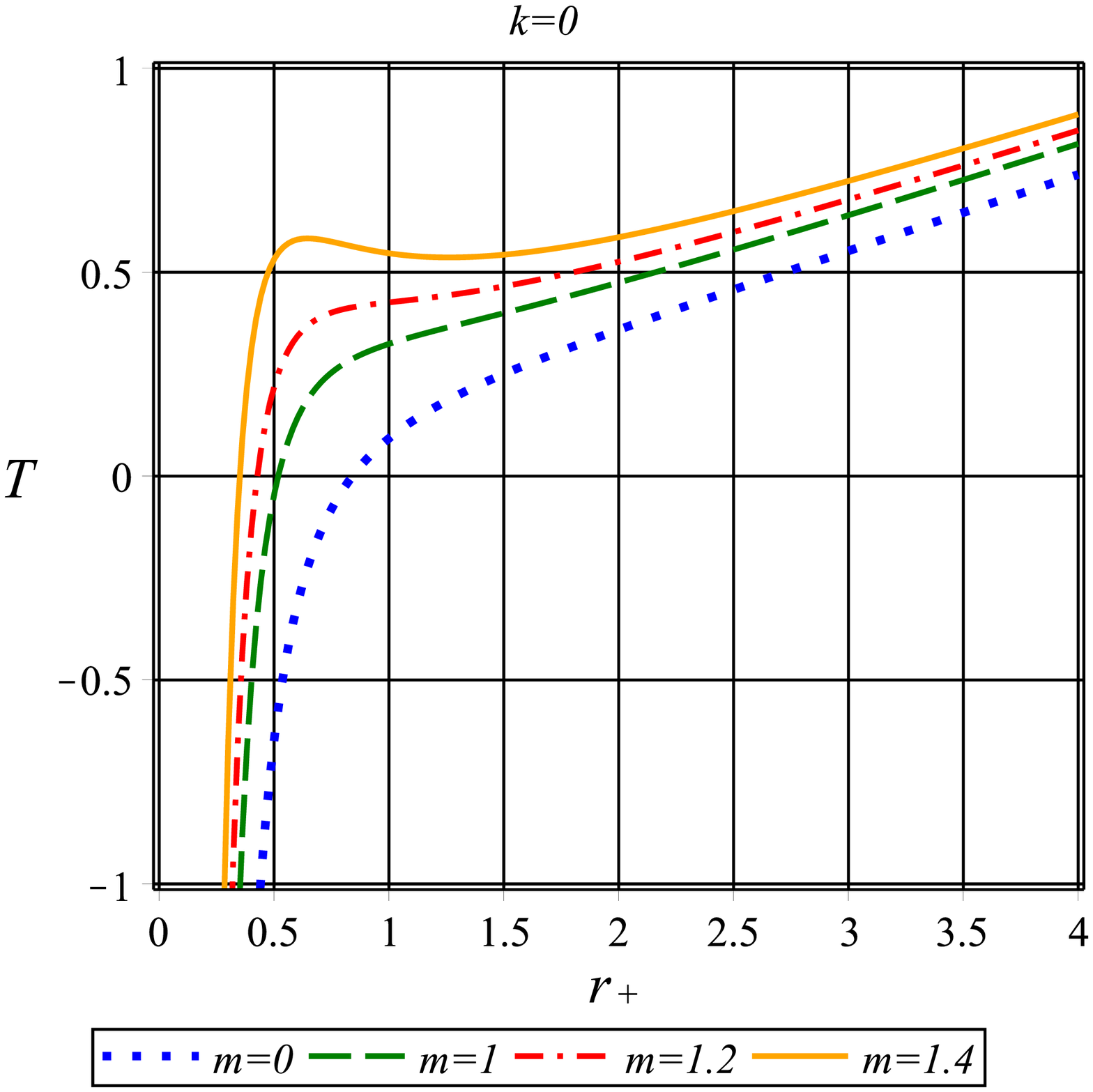}\includegraphics[width=50 mm]{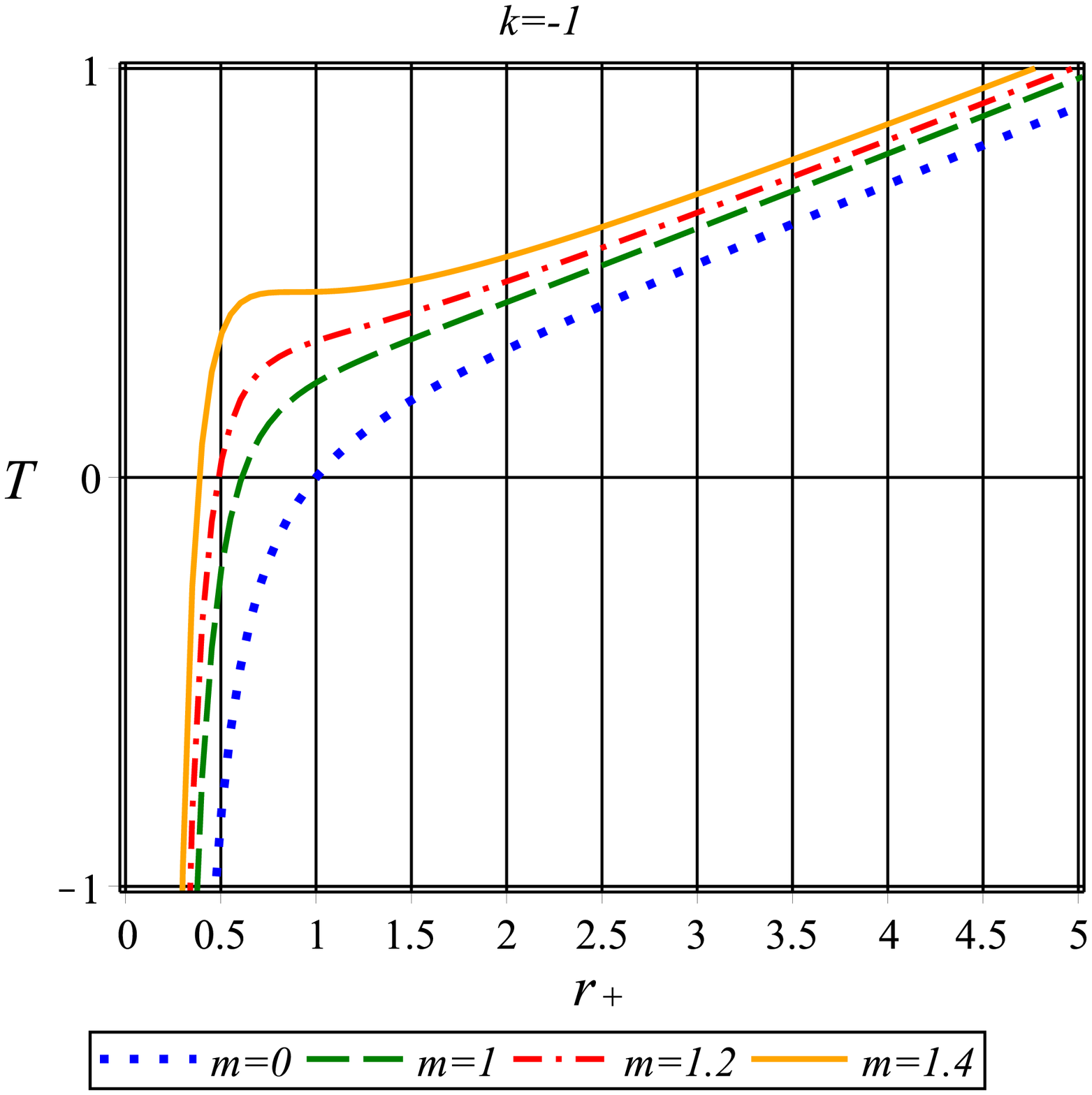} &  &  &
\end{array}%
$%
\end{center}
\caption{Temperature of the second model (Eq. (\protect\ref{eq43})) in terms
of horizon radius, for $\protect\xi=1$ and $\protect\varepsilon=1$ (unit
value for other parameters).}
\label{figT-2}
\end{figure}

Now, we can plot the radius of the horizon. In Fig. \ref{figT-2}, we plot
the temperature for various values of $m$ and obtain result similar to the
ones obtained for the model motivated from loop quantum gravity.
Interestingly, there is a critical horizon radius for which the temperature
is constant (for example, see solid orange line of Fig. \ref{figT-2} for $%
k=-1$).

\begin{figure}[h!]
\begin{center}
$%
\begin{array}{cccc}
\includegraphics[width=50 mm]{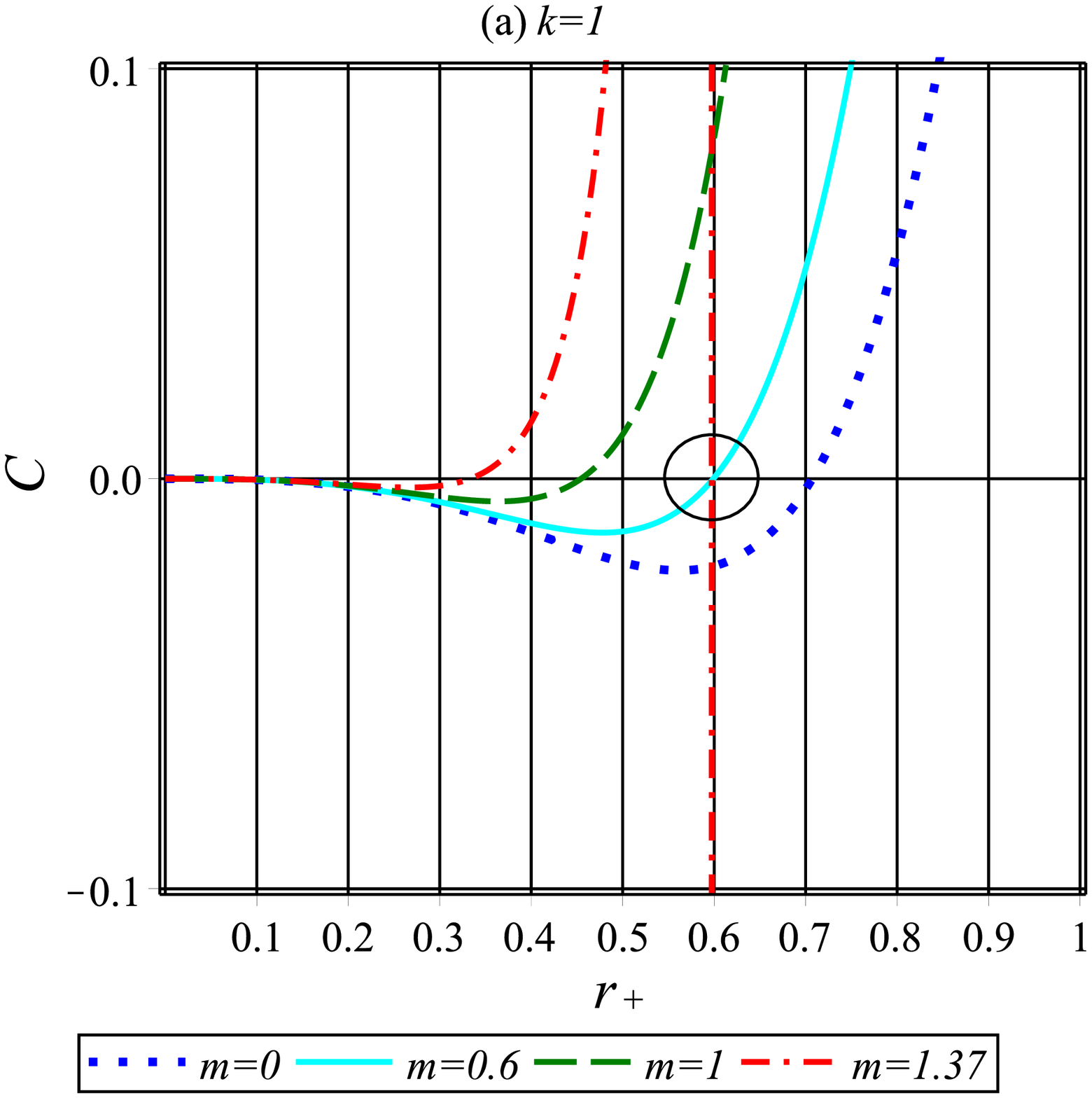}%
\includegraphics[width=50
mm]{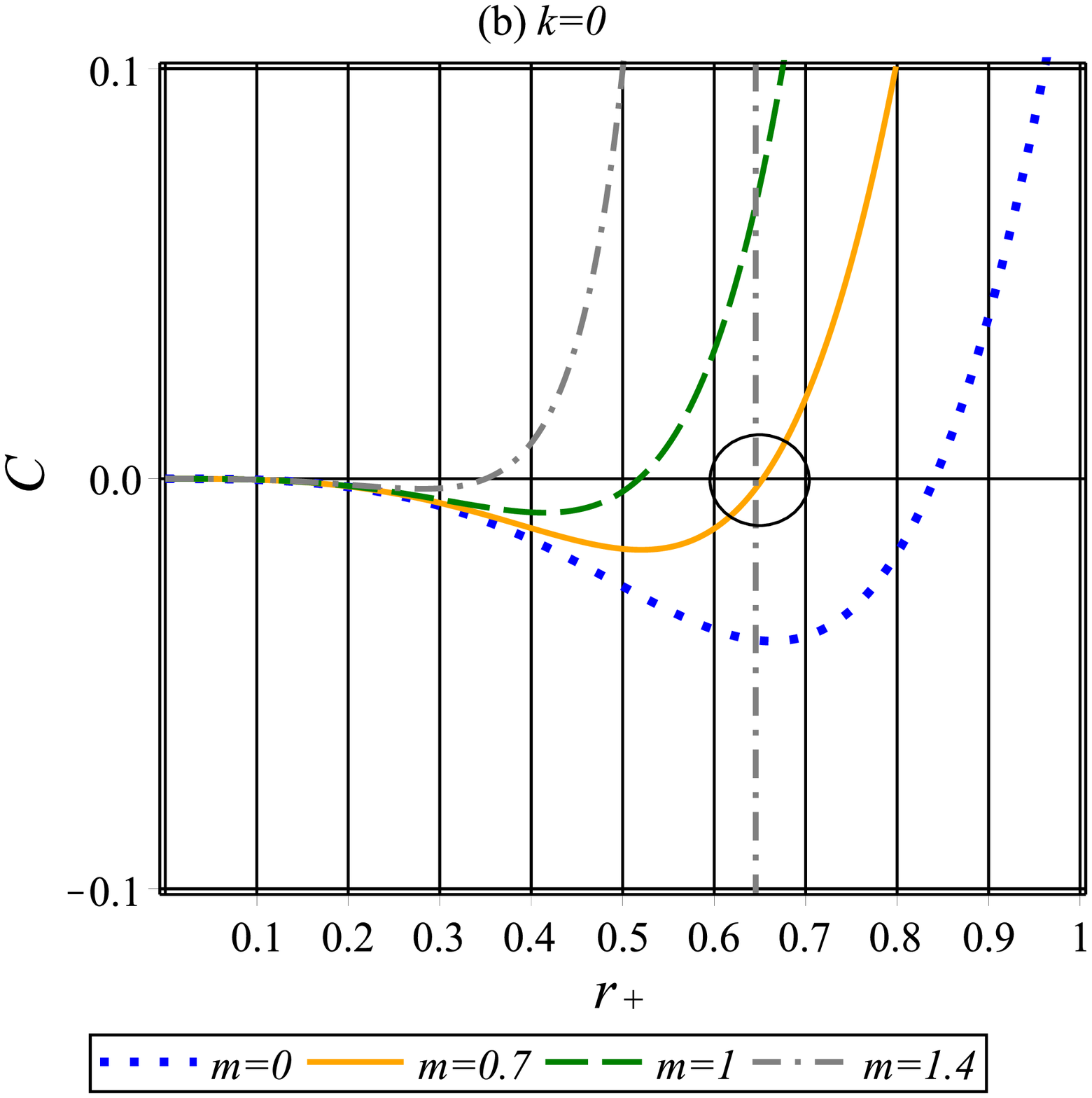}\includegraphics[width=50 mm]{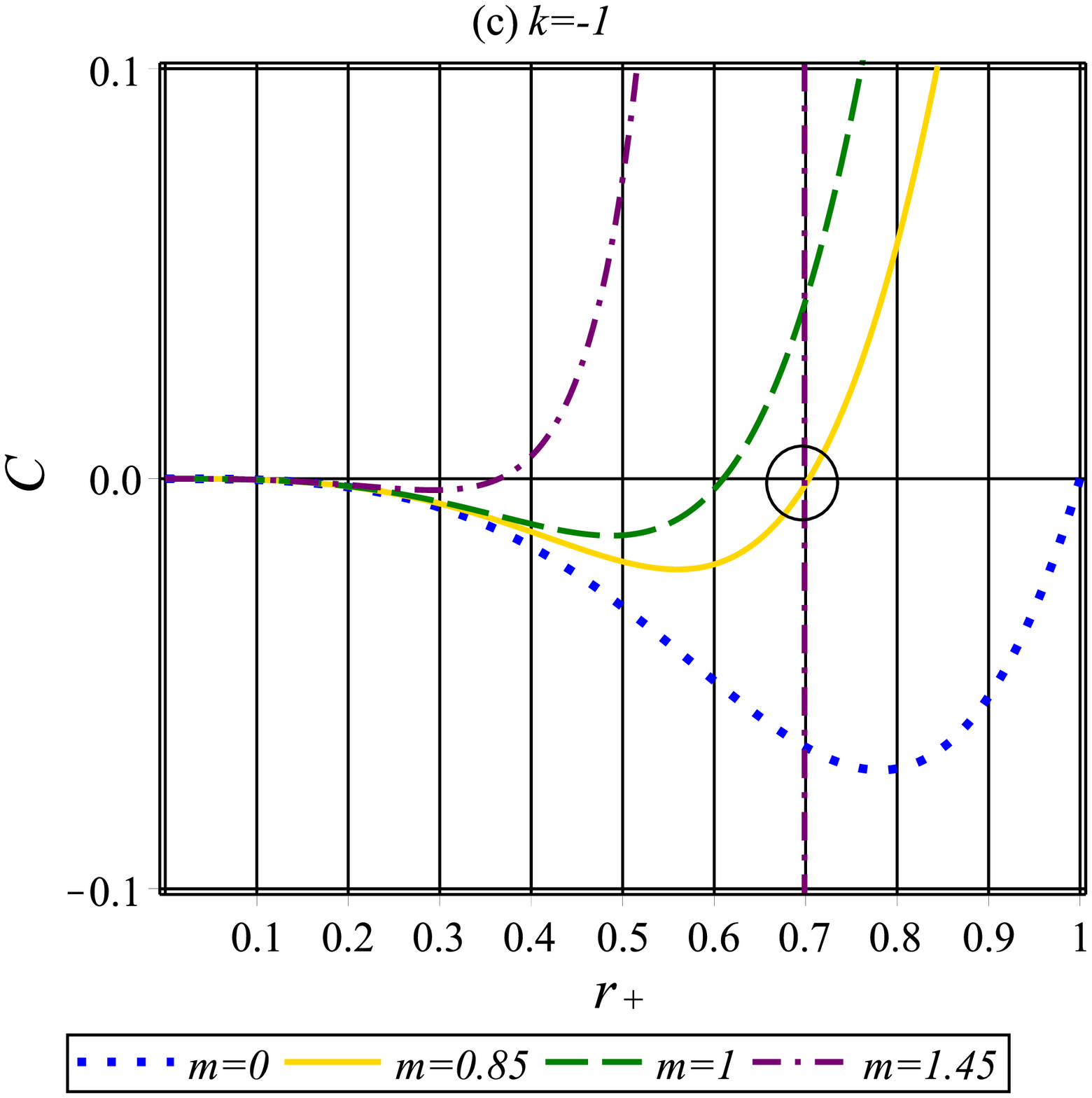} &  &  &
\end{array}%
$%
\end{center}
\caption{Specific heat of the second model (Eq. (\protect\ref{eq45})) in
terms of $r_{+}$, with unit values for the model parameters.}
\label{figC-2}
\end{figure}

Then, the black hole mass can be obtained as
\begin{eqnarray}  \label{eq44}
M=\frac{3}{16\pi}\left[\frac{\xi\varepsilon k}{e^{\xi\varepsilon}-1}%
r_{+}^{2} +\frac{\xi\varepsilon}{e^{\xi\varepsilon}-1}\left[\frac{r_{+}^{4}}{%
l^{2}}-2e^{2}\ln\left(r_{+}\right) +m^{2}\left(\frac{c_{0}c_{1}}{3}%
r_{+}^{3}+c_{0}^{2}c_{2}r_{+}^{2}+2c_{0}^{3}c_{3}r_{+}\right)\right]\right].
\end{eqnarray}
The specific heat is given by
\begin{eqnarray}  \label{eq45}
C=\frac{3r_{+}^{2}}{4}\frac{\left[\frac{2k}{r_{+}}\frac{\xi\varepsilon}{%
e^{\xi\varepsilon}-1} +\frac{\xi\varepsilon}{e^{\xi\varepsilon}-1}\left[%
\frac{4r_{+}}{l^{2}}-\frac{2e^{2}}{r_{+}^{3}} +m^{2}\left(c_{0}c_{1}+2\frac{%
c_{0}^{2}c_{2}}{r_{+}} +2\frac{c_{0}^{3}c_{3}}{r_{+}^{2}}\right)\right]%
\right]}{\left[-\frac{2k}{r_{+}^{2}}\frac{\xi\varepsilon}{%
e^{\xi\varepsilon}-1} +\frac{\xi\varepsilon}{e^{\xi\varepsilon}-1}\left[%
\frac{4}{l^{2}}+\frac{6e^{2}}{r_{+}^{4}} +m^{2}\left(-2\frac{c_{0}^{2}c_{2}}{%
r_{+}^{2}}-4\frac{c_{0}^{3}c_{3}}{r_{+}^{3}}\right)\right]\right]}
\end{eqnarray}
According to Fig. \ref{figC-2}, it is evident that the smaller values $m$
produce the negative value of specific heat. On the other hand, larger
values of $m$ produce a phase transition from a stable to an unstable phase.
So, for the thermodynamically stable model, we should have $%
m_{min}<m<m_{max} $. For the unit value of the model parameters, we find $%
m_{min}=0.6$ and $m_{max}=1.37$ for $k=1$ (see Fig. \ref{figC-2} (a)), $%
m_{min}\approx0.7$ and $m_{max}\approx1.4$ for $k=0$ (see Fig. \ref{figC-2}
(b)). We also have $m_{min}\approx0.85$ and $m_{max}\approx1.45$ for $k=-1$
(see Fig. \ref{figC-2} (c)). These plots show that the phase transition is
possible for the second model.

\subsection{Horizon Problem}

\label{sec3-4} It has been proposed that the horizon problem can be resolved
with suitable rainbow functions \cite{MS, h},
\begin{eqnarray}  \label{eq46}
f\left(\varepsilon\right)=g\left(\varepsilon\right)=\frac{1}{%
1-\lambda\varepsilon},
\end{eqnarray}
where $\xi$ is a dimensionless parameters of the order of unity. In this
case, the entropy and temperature of the system can be written as
\begin{equation}  \label{eq47}
S=\frac{\left(1-\lambda\varepsilon\right)^{3}r_{+}^{3}}{4}.
\end{equation}
and
\begin{equation}  \label{eq48}
T_{H}=\frac{1}{4\pi}\left[\frac{2k}{r_{+}}+\left(1-\lambda\varepsilon%
\right)^{2}\left[\frac{4r_{+}}{l^{2}} -\frac{2e^{2}}{r_{+}^{3}}%
+m^{2}\left(c_{0}c_{1}+2\frac{c_{0}^{2}c_{2}}{r_{+}}+2\frac{c_{0}^{3}c_{3}}{%
r_{+}^{2}}\right)\right]\right].
\end{equation}

\begin{figure}[h!]
\begin{center}
$%
\begin{array}{cccc}
\includegraphics[width=55 mm]{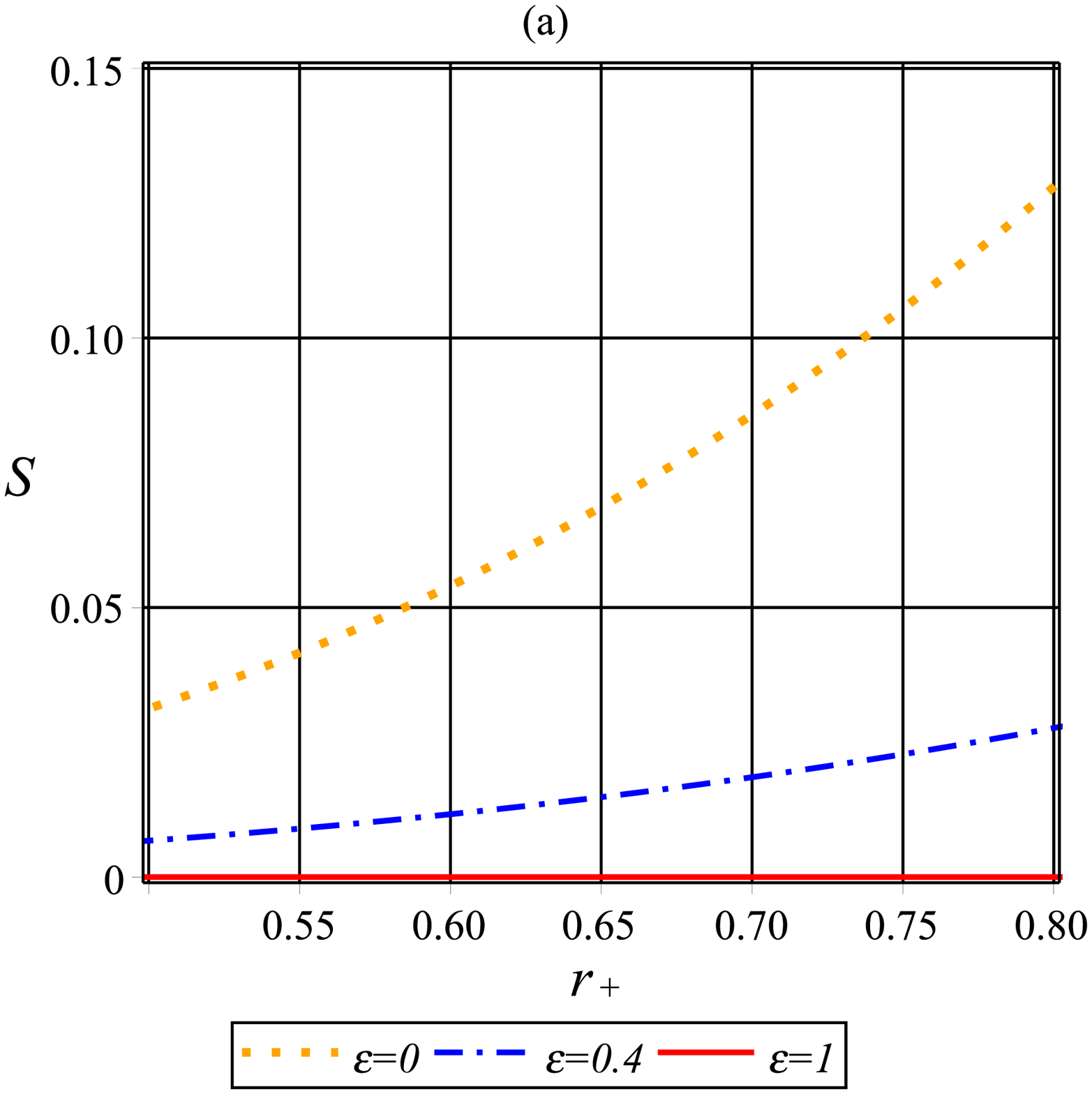}%
\includegraphics[width=55
mm]{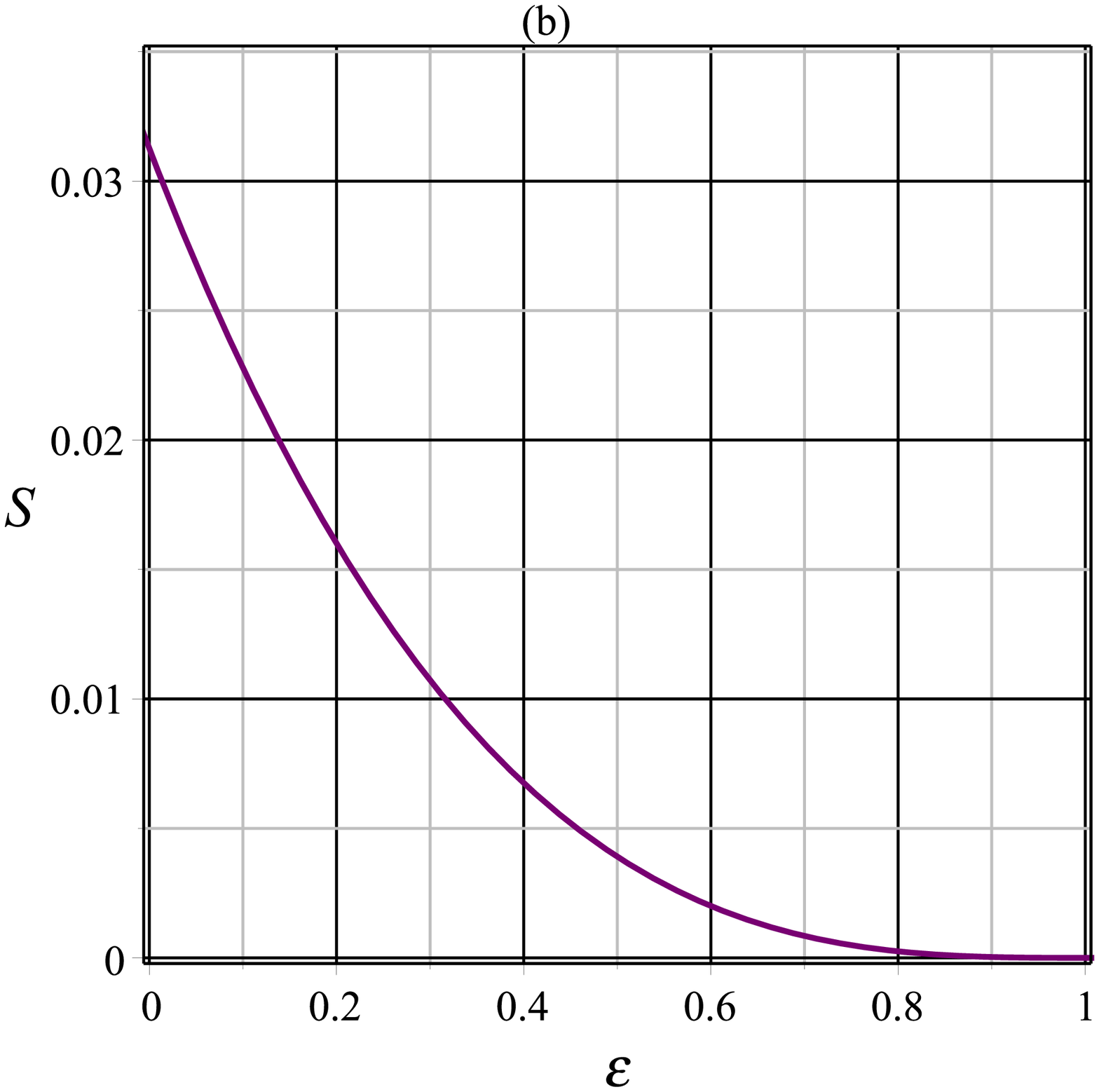} &  &  &
\end{array}%
$%
\end{center}
\caption{Typical behavior of the entropy in the third model (equation (%
\protect\ref{eq47})) for $\protect\lambda=1$. (a) in terms of $r_{+}$ for
different values of $\protect\varepsilon$, and (b) in terms of $\protect%
\varepsilon$ for $r_{+}=0.5$. Allowed region are separated by orange box.}
\label{figS-3}
\end{figure}

\begin{figure}[h!]
\begin{center}
$%
\begin{array}{cccc}
\includegraphics[width=60 mm]{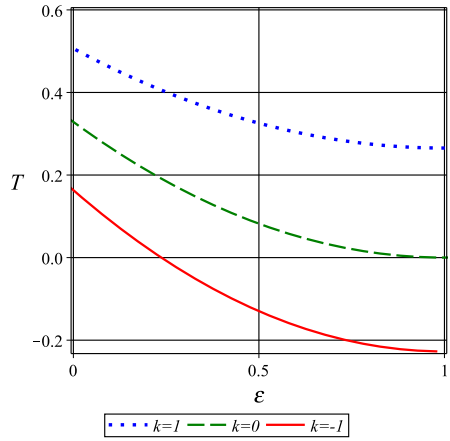} &  &  &
\end{array}%
$%
\end{center}
\caption{Typical behavior of the temperature in the third model (Eq. (%
\protect\ref{eq48})), for unit value for all model parameters.}
\label{figT-3}
\end{figure}

In order to have well defined model (positive entropy and temperature), we
should have $\varepsilon\leq\frac{1}{\lambda}$ and $m>m_{min}$. In the plots
of Fig. \ref{figS-3}, we observe the behavior of the entropy. Now Fig. \ref%
{figS-3} (a) demonstrates that there is an upper limit for the energy, below
which the entropy is negative. For the selected value $\lambda=1$, we
observe that $\varepsilon_{max}=1$. Here $S=0$, and $S\geq0$ for $%
\varepsilon\leq\varepsilon_{max}$. It should be noted that general behavior
is similar for $k=0$, and $k= \pm1$. It is also illustrated by Fig. \ref%
{figS-3} (b) which plots the behavior of the entropy with $\varepsilon$.%
\newline
In Fig. \ref{figT-3}, we can verify our previous results. According to Fig. %
\ref{figT-3}, we shall denote the maximum $\varepsilon=\frac{1}{\lambda}$ by
$\varepsilon_{max}$ ($\varepsilon=1$ in plot), Here the value of temperature
does not depend on $m$. For $k=0$ and $k=1$ temperature is positive for
suitable mass. However, for $k=-1$, value of temperature is negative at this
energy. Hence, we find that $T_{H}$ is positive for $\varepsilon<%
\varepsilon_{max}$. The positive temperature occurs, when $%
\varepsilon>\varepsilon_{max}$ is not allowed. So, both $\varepsilon$ and $m$
are constrained as $\varepsilon<\varepsilon_{max}$ and $m>m_{min}$.\newline

\begin{figure}[h!]
\begin{center}
$%
\begin{array}{cccc}
\includegraphics[width=70 mm]{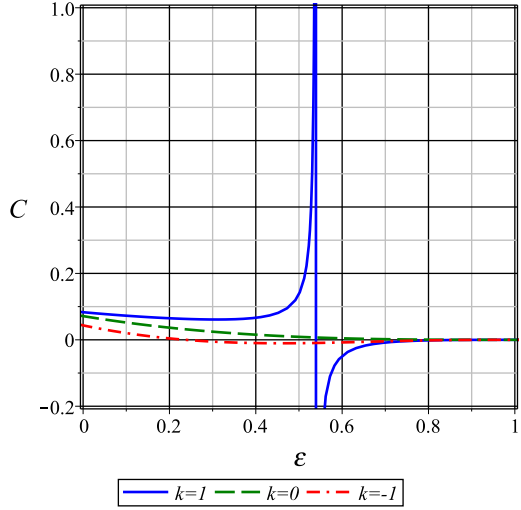} &  &  &
\end{array}%
$%
\end{center}
\caption{Typical behavior of the specific heat in terms of $\protect%
\varepsilon$ in the third model (Eq. (\protect\ref{eq50})), with unit value
for all model parameters.}
\label{figT-3B}
\end{figure}

Then, we can find,
\begin{eqnarray}  \label{eq49}
M&=&\frac{3kr_{+}^{2}\left(1-\lambda\varepsilon\right)^{3}}{16\pi}  \notag \\
&+&\frac{3\left(1-\lambda\varepsilon\right)^{5}}{16\pi} \left[\frac{r_{+}^{4}%
}{l^{2}}-2e^{2}\ln\left(r_{+}\right)+m^{2}\left(\frac{c_{0}c_{1}}{3}%
r_{+}^{3}+c_{0}^{2}c_{2}r_{+}^{2}+2c_{0}^{3}c_{3}r_{+}\right)\right]. \\
\end{eqnarray}
\begin{eqnarray}  \label{eq50}
C&=&\frac{3\left(1-\lambda\varepsilon\right)^{3}r_{+}^{2}}{4}\frac{\left[%
\frac{2k}{r_{+}} +\left(1-\lambda\varepsilon\right)^{2}\left[\frac{4r_{+}}{%
l^{2}}-\frac{2e^{2}}{r_{+}^{3}} +m^{2}\left(c_{0}c_{1}+2\frac{c_{0}^{2}c_{2}%
}{r_{+}}+2\frac{c_{0}^{3}c_{3}}{r_{+}^{2}}\right)\right]\right]}{\left[-%
\frac{2k}{r_{+}^{2}} +\left(1-\lambda\varepsilon\right)^{2}\left[\frac{4}{%
l^{2}}+\frac{6e^{2}}{r_{+}^{4}} +m^{2}\left(-2\frac{c_{0}^{2}c_{2}}{r_{+}^{2}%
}-4\frac{c_{0}^{3}c_{3}}{r_{+}^{3}}\right)\right]\right]}.
\end{eqnarray}
Graphical analysis of specific heat is represented in Fig. \ref{figT-3B}. It
shows the variation of the specific heat with $\varepsilon$. For $k=1$, we
can see the first and second phase transitions. We confirm the previous
result, as $\varepsilon<\varepsilon_{max}$ is crucial to have well defined
model.

%%%%%%%%%%%%%%%%%%%%%%%%%%%%%%%%%%%

\section{Criticality in the Extended Phase Space}

\label{sec44} Now, we give a discussion of the critical behavior of the
Yang-Mills black hole solution in the using the extended phase space \cite%
{exten1, exten2}. In the extended phase space, the cosmological constant is
identified with a thermodynamic pressure as \cite{crit14, crit16}
\begin{equation}
P=-\frac{\Lambda }{8\pi }=\frac{3}{4\pi l^{2}}.  \label{pressure}
\end{equation}

Substituting the pressure from Eq. (\ref{pressure}) in Eq. (\ref{Temp}), one
can obtain the following equation of state
\begin{equation}
P=\frac{3}{8\pi }\left( \frac{e^{2}}{r_{+}^{4}}-\frac{m^{2}c_{0}^{3}c_{3}}{%
r_{+}^{3}}-\frac{\left[ m^{2}c_{0}^{2}c_{2}+kg^{2}(\varepsilon )\right] }{%
r_{+}^{2}}+\frac{\left[ 4\pi f(\varepsilon )g(\varepsilon )T-m^{2}c_{0}c_{1}%
\right] }{2r_{+}}\right) .  \label{EoS}
\end{equation}

Due to the fact that in $(4+1)-$dimensions, the specific volume ($v$) is
related to the event horizon radius as $v=\frac{{4{r_{+}}\ell _{\mathrm{P}}^{%
{3}}}}{{3}}$, we can work with $P=P(T,r_{+})$, instead of $P=P(T,v)$, as it
will produce the same thermodynamic behavior. In other words, the
criticality, phase transition and, in general, the behavior of $P-v$ diagram
is equivalent to $P-r_{+}$ diagram. Regarding Eq. (\ref{EoS}), we observe
that it is reasonable to define an effective (shifted) temperature $T_{eff}$%
, and horizon topology factor $k_{eff}$ as
\begin{eqnarray}
T_{eff} &=&T-\frac{m^{2}c_{0}c_{1}}{4\pi f(\varepsilon )g(\varepsilon )},
\label{Teff} \\
k_{eff} &=&k+\frac{m^{2}c_{0}^{2}c_{2}}{g^{2}(\varepsilon )}.  \label{Keff}
\end{eqnarray}

In order to obtain the critical point of isothermal $P-r_{+}$ diagram, we
use the inflection point property of such a diagram as
\begin{equation}
\left( \frac{\partial P}{\partial r_{+}}\right) _{T_{eff}}=0,~~~~~\left(
\frac{\partial ^{2}P}{\partial r_{+}^{2}}\right) _{T_{eff}}=0.
\label{Inflection}
\end{equation}

After some simplification, we find the following expression corresponding to
Eq. (\ref{Inflection})
\begin{eqnarray}
\left( \frac{\partial P}{\partial r_{+}}\right) _{T_{eff}} &=&\frac{-3}{8\pi
{r}_{+}^{5}}\left[ 2\pi f(\varepsilon )g(\varepsilon )T_{eff}{r}%
_{+}^{3}-2g^{2}(\varepsilon ){k_{eff}r}_{+}^{2}-3\,{{c_{0}}}^{3}{c_{3}}\,{m}%
^{2}r_{+}+4{e}^{2}\right] =0  \label{dP} \\
\left( \frac{\partial ^{2}P}{\partial r_{+}^{2}}\right) _{T_{eff}} &=&\frac{3%
}{4\pi \,{r}_{+}^{6}}\left[ 2\pi f(\varepsilon )g(\varepsilon )T_{eff}{r}%
_{+}^{3}-3g^{2}(\varepsilon ){k_{eff}r}_{+}^{2}-6{{c_{0}}}^{3}{c_{3}}\,{m}%
^{2}{r}_{+}+10{e}^{2}\right] =0.  \label{ddP}
\end{eqnarray}%
The critical quantities are obtained by solving Eqs. (\ref{dP}) and (\ref%
{ddP}), simultaneously
\begin{eqnarray}
&&r_{c}=-\frac{3m^{2}c_{0}^{3}c_{3}\pm \Theta }{2k_{eff}g^{2}(\varepsilon )},
\label{criticalQ} \\
&&T_{eff}|_{c}=\frac{2k_{eff}^{2}g^{3}(\varepsilon )\left[ \Theta ^{2}\mp
3m^{2}c_{0}^{3}c_{3}\Theta -8e^{2}k_{eff}g^{2}(\varepsilon )\right] }{\pi
f(\varepsilon )(-3m^{2}c_{0}^{3}c_{3}\pm \Theta )^{3}},  \notag \\
&&P_{c}=\frac{3k_{eff}^{3}g^{6}(\varepsilon )\left[ \Theta ^{2}\mp
2m^{2}c_{0}^{3}c_{3}\Theta -12e^{2}k_{eff}g^{2}(\varepsilon
)-3m^{4}c_{0}^{6}c_{3}^{2}\right] }{2\pi (-3m^{2}c_{0}^{3}c_{3}\pm \Theta
)^{4}},  \notag
\end{eqnarray}%
where $\Theta =\sqrt{9m^{4}c_{0}^{6}c_{3}^{2}+24e^{2}k_{eff}g^{2}(%
\varepsilon )}$. The two branches of critical quantities distinguishing by
the sign behind $\Theta $, where for the lower sign, there is results are
not physical, for any real positive values of the critical quantities. So,
we will only analyze this system with the upper sign.

\begin{figure}[h!]
\begin{center}
$%
\begin{array}{cc}
\includegraphics[width=75 mm]{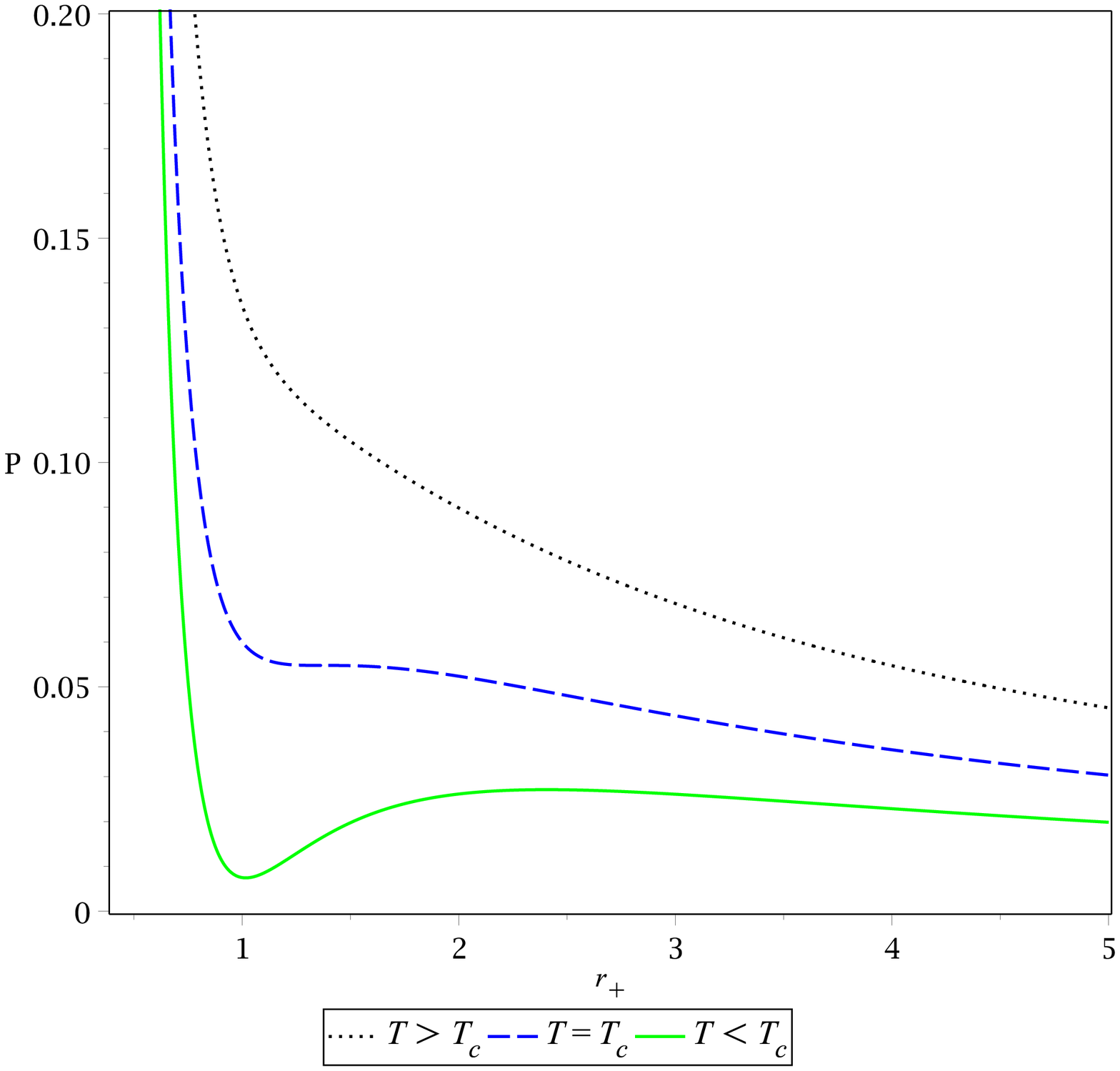} & \includegraphics[width=75 mm]{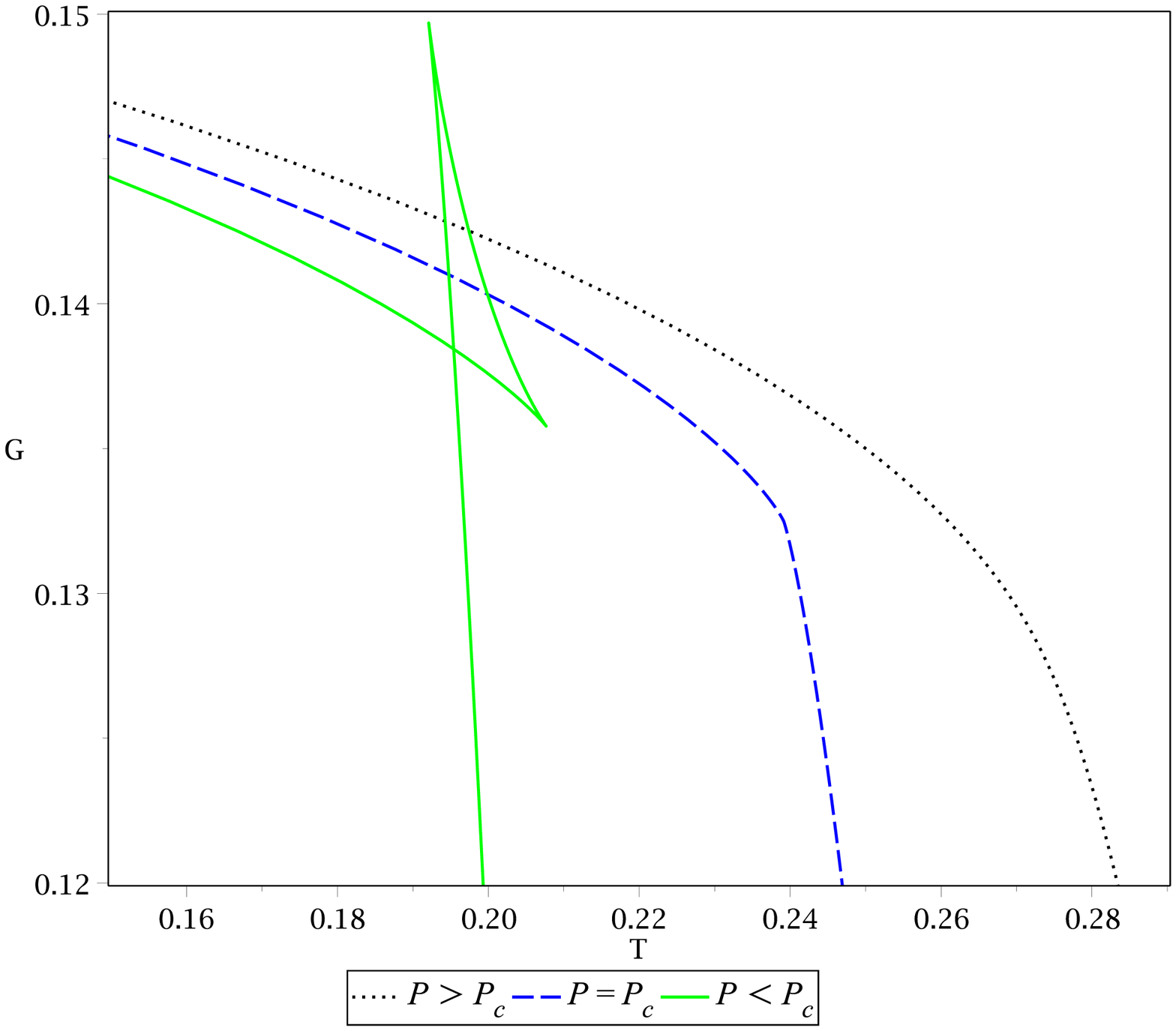}%
\end{array}%
$%
\end{center}
\caption{$P-r_+$ (left) and $G-T$ (right) diagrams for $%
e=m=c_0=c_1=c_2=c_3=g(\protect\varepsilon)=1$ and $k_{eff}=1$ ($k=1$ with $%
c_{2}=0$ or $k=0$ with $c_{2}=1$ or $k=-1$ with $c_{2}=2$). The blue dashed
line corresponds to the critical temperature (left) and the critical
pressure (right).}
\label{FigPVGT}
\end{figure}

Although the Van der Waals phase transition and critical behavior are
observed only for spherical horizon topology in the Einstein-AdS gravity,
here in the massive gravity scenario, we can build such behavior for all
topologies. As we indicate in the caption of Fig. \ref{FigPVGT}, it is
obvious that by adjusting the massive parameters (or rainbow function), one
can find the Van der Waals like behavior.

To confirm the results, we can study the Gibbs free energy per unit volume $%
\omega _{k}$ as follows
\begin{equation}
G=M-TS=\frac{6e^{2}+3k_{eff}g^{2}(\varepsilon
)r_{+}^{2}+12m^{2}c_{0}^{3}c_{3}r_{+}-4\pi Pr_{+}^{4}-18e^{2}\ln r_{+}}{%
48\pi f(\varepsilon )g^{4}(\varepsilon )}.  \label{Gibbs}
\end{equation}

According to the right panel of Fig. \ref{FigPVGT}, we observe a first-order
phase transition for different topologies, which is characterized by the
swallow-tail shape of the Gibbs free energy for $P<P_{c}$.

%%%%%%%%%%%%%%%%%%%%%%%%%%%%%%%%%%%%%%%%%%%%%%%%%%%%%%%%%%%%%%%%%%%%%

%%%%%%%%%%%%%%%%%%%%%%%%%%%%%%%%%%%%%%%%%%%%%%%%%%%%%%%%%%%%%%%%%%%%%

%%%%%%%%%%%%%%%%%%%%%%%%%%%%%%%%%%%%%%%%%%%%%%%%%%%%%%%%%%%%%%%%%%%%%%%%%%%%%%%%%%%%%%%%%%%%%%%

\section{Thermal Fluctuations}

\label{sec4} %%%%%%%%%%%%%%%%%%%%%%%%%%%%%%%%%%%

It has been argued that the thermodynamics of black hole should be corrected
due to thermal fluctuations \cite{T1}. Such fluctuations can be analyzed
using the partition function for such a system. As the AdS black holes are
dual to conformal field theories, it is possible to analyze the fluctuations
to the black hole thermodynamics using the statistical mechanical partition
function of microstates, which can be obtained from the dual conformal field
theory \cite{T2}. It is possible to explicitly write down such a partition
function as
\begin{eqnarray}  \label{Z}
Z=\int_{0}^{\infty}{\Omega e^{-\frac{E}{T}}dE},
\end{eqnarray}
where $\Omega$ denotes the density of state in canonical ensemble, which is
proportional to
\begin{eqnarray}  \label{Omega}
\Omega\propto\int{\frac{e^{s}}{T^{2}}dT}.
\end{eqnarray}
Here $s$ denotes the exact (corrected) entropy. Applying the inverse Laplace
transformation to Eq. (\ref{Z}) yields
\begin{eqnarray}  \label{Omega2}
\Omega\propto\int{\frac{z e^{\frac{E}{T}}}{T^{2}}dT}.
\end{eqnarray}
Combining Eqs. (\ref{Omega}) and (\ref{Omega2}) yields
\begin{equation}  \label{s1}
s=\ln{Z}+\frac{E}{T}.
\end{equation}
This is identical to the statistical relation,
\begin{equation}  \label{s2}
s=\ln{\Omega}.
\end{equation}
If we assume $S$ to be the equilibrium entropy, and use Taylor expansion of $%
s$ in Eq. (\ref{Omega2}), then after some calculations, we obtain \cite{T1,
T2}
\begin{equation}  \label{s}
s=S-\frac{\alpha}{2}\ln\left[\frac{S^{\prime\prime}T^{\prime}-S^{\prime}T^{%
\prime\prime}}{(T^{\prime})^{3}}\right]+\cdots,
\end{equation}
where prime denotes derivative with respect to the horizon radius $r_{+}$,
ie., $S^{\prime}=\frac{dS}{dr_{+}}$. Also, the constant $\alpha$ is added by
hand to track correction terms \cite{EPL}. Here, $\alpha=0$ reproduces
results in the absence of such logarithmic corrections, and $\alpha=1$
produces logarithmic corrections. In Eq. (\ref{s}), we have neglected
higher-order terms in the Taylor expansion (which produce higher-order
corrections). Hence, the first-order correction occurs in form of a
logarithmic correction term to the entropy \cite{EPJC1, PRD1}.\newline
By using the general temperature and general entropy, one can obtain this
logarithmic corrected entropy as
\begin{equation}  \label{s-general}
s=\frac{r_{+}^{3}}{4g^{3}\left(\varepsilon\right)}-\frac{\alpha}{2}\ln\left[%
\frac{15\pi^{2}r_{+}^{9}}{4} \frac{f^{2}\left(\varepsilon\right)}{%
g\left(\varepsilon\right)} \frac{(m^{2}+kg^{2}\left(\varepsilon\right))\frac{%
2r_{+}^{2}}{5}+m^{2}r_{+}-\frac{1}{5}(9+2r_{+}^{4})} {((m^{2}+kg^{2}\left(%
\varepsilon\right))\frac{r_{+}^{2}}{2}+m^{2}r_{+}-r_{+}^{4}-\frac{3}{2})^{3}}%
\right],
\end{equation}
where, we have assumed $c_{0}=c_{1}=c_{2}=c_{3}=e=1$ for simplicity. Then,
we can obtain corrected mass via
\begin{eqnarray}  \label{M}
M_{c}=\int Tds.
\end{eqnarray}
To see effect of the logarithmic correction, we plot the entropy in Fig. \ref%
{s-log}. We observe that the thermal fluctuations are important in smaller $%
r_{+}$. It has been argued that when the black hole size reduced due to the
Hawking radiation, the thermal fluctuations become important \cite{NPB}. As
general behavior does not depend on $\varepsilon$, we fix their values to
study thermodynamics behaviors generally. We show that thermal fluctuations
produce the negative entropy for larger $r_{+}$. Depend on the model ($%
f\left(\varepsilon\right))$, black hole in presence of thermal fluctuations
may be stable or unstable.\newline

\begin{figure}[h!]
\begin{center}
$%
\begin{array}{cccc}
\includegraphics[width=50 mm]{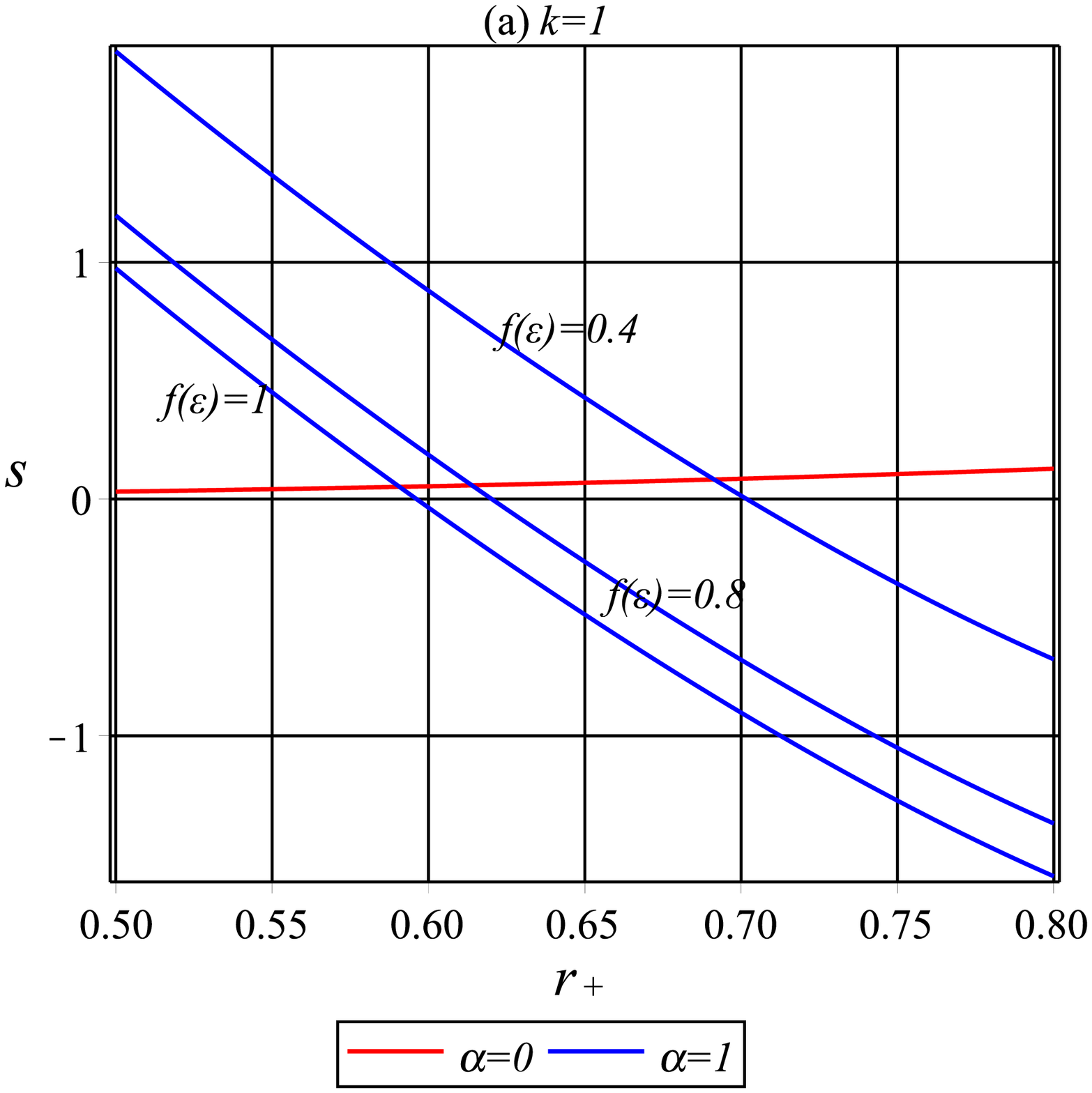}%
\includegraphics[width=50
mm]{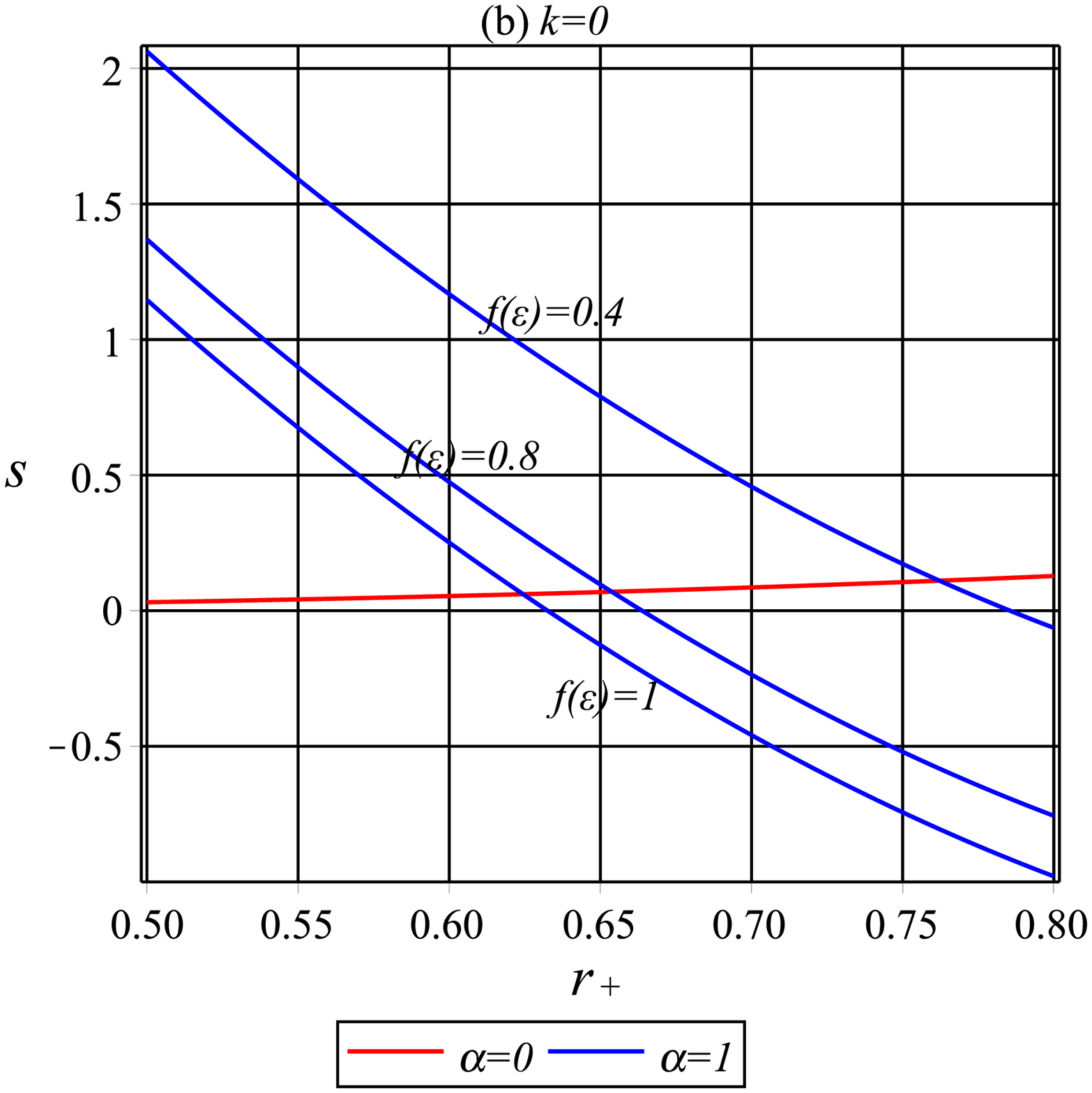}\includegraphics[width=50 mm]{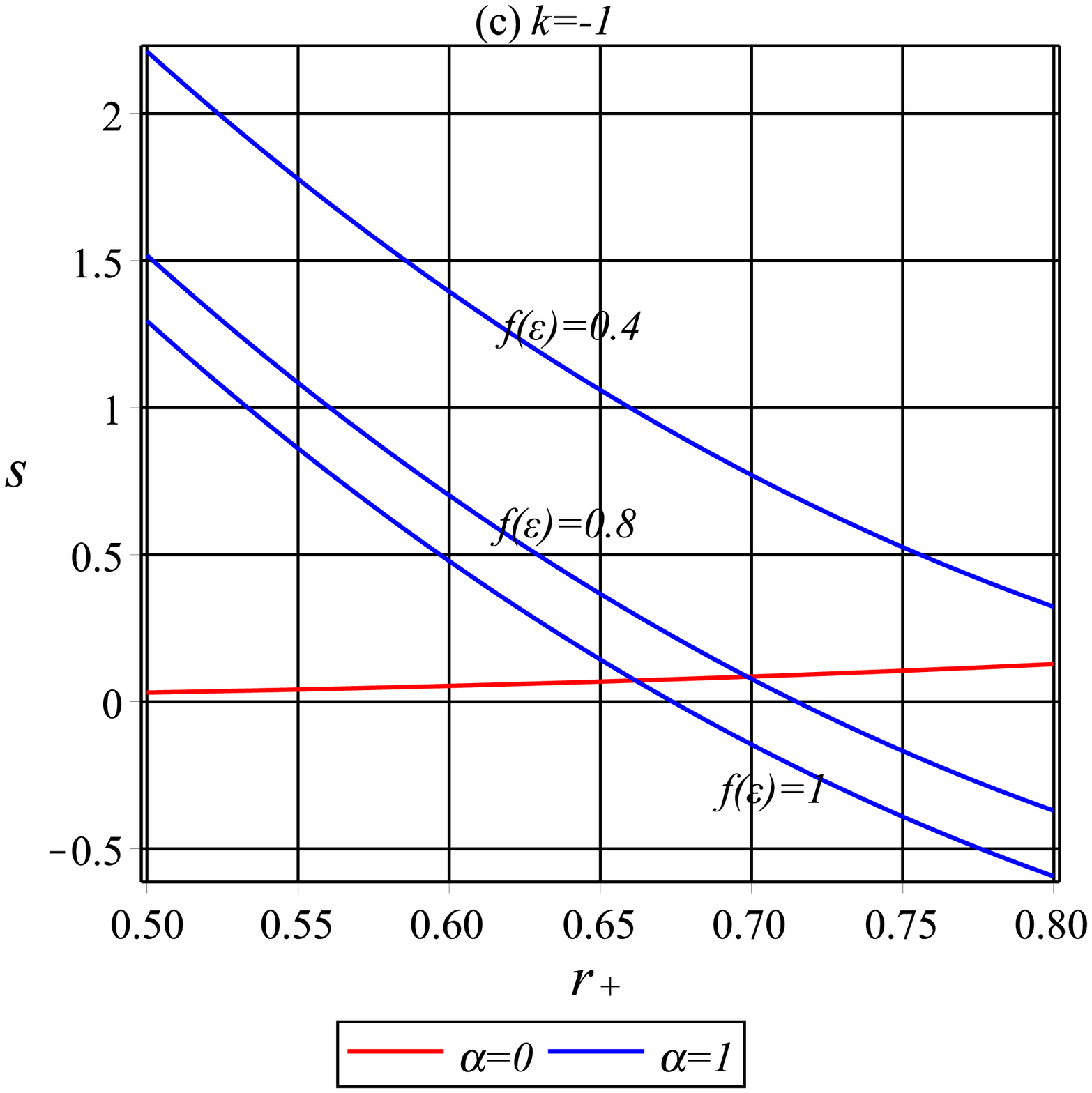} &  &  &
\end{array}%
$%
\end{center}
\caption{Logarithmic corrected entropy in terms of $r_{+}$ for $g\left(%
\protect\varepsilon\right)\approx1$, and unit value for other model
parameters.}
\label{s-log}
\end{figure}

\begin{figure}[h!]
\begin{center}
$%
\begin{array}{cccc}
\includegraphics[width=50 mm]{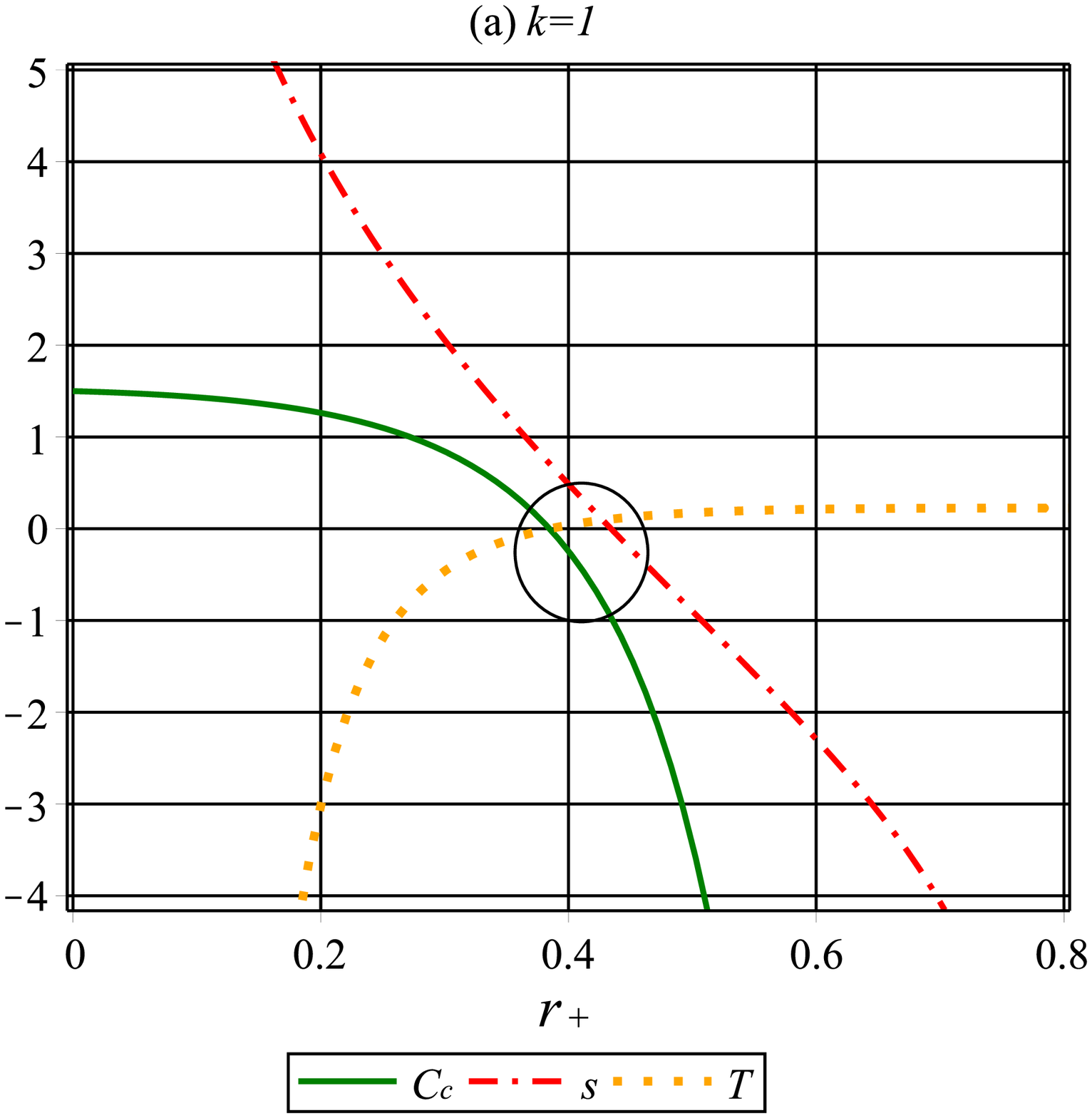}%
\includegraphics[width=50
mm]{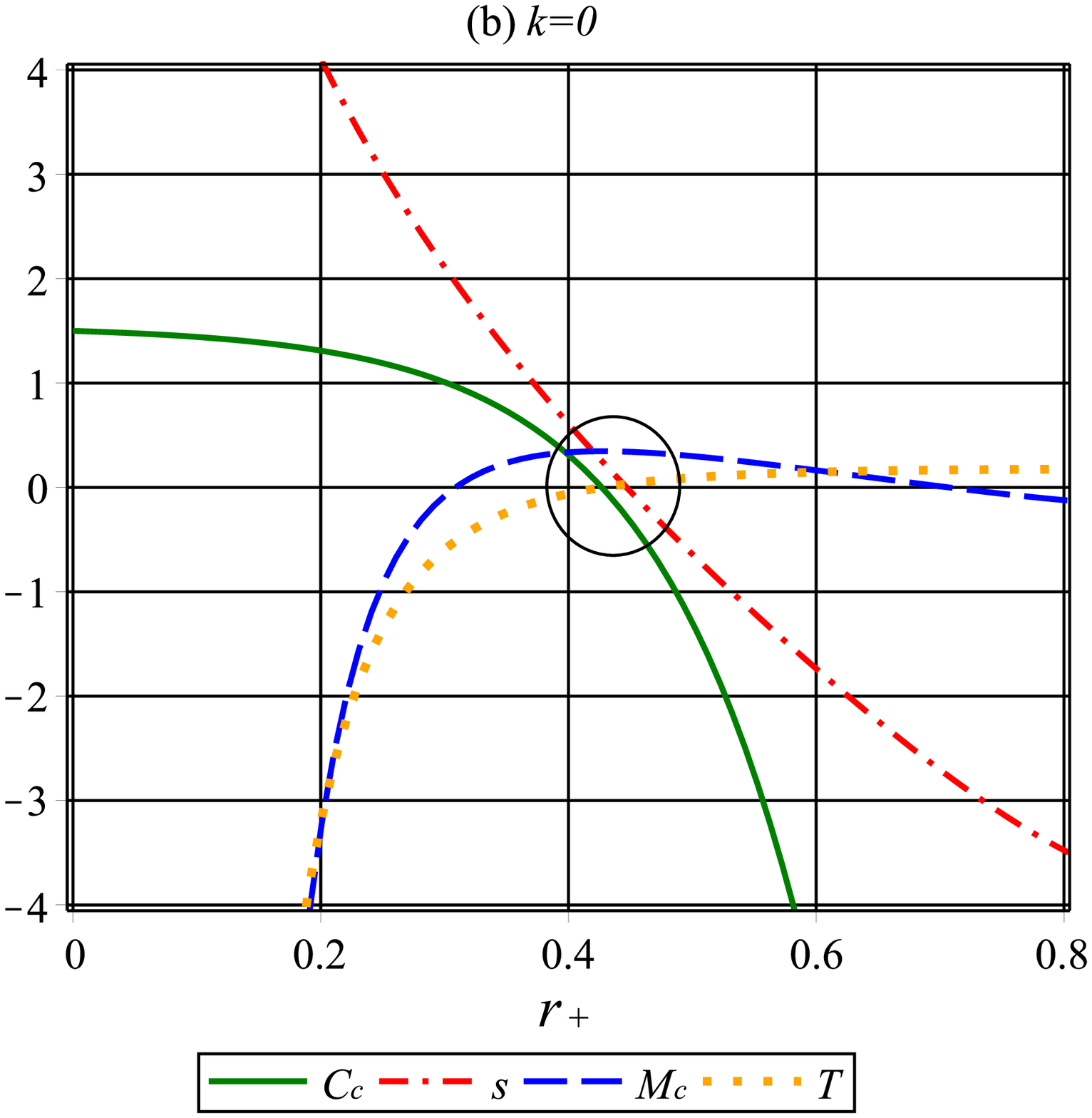}\includegraphics[width=50 mm]{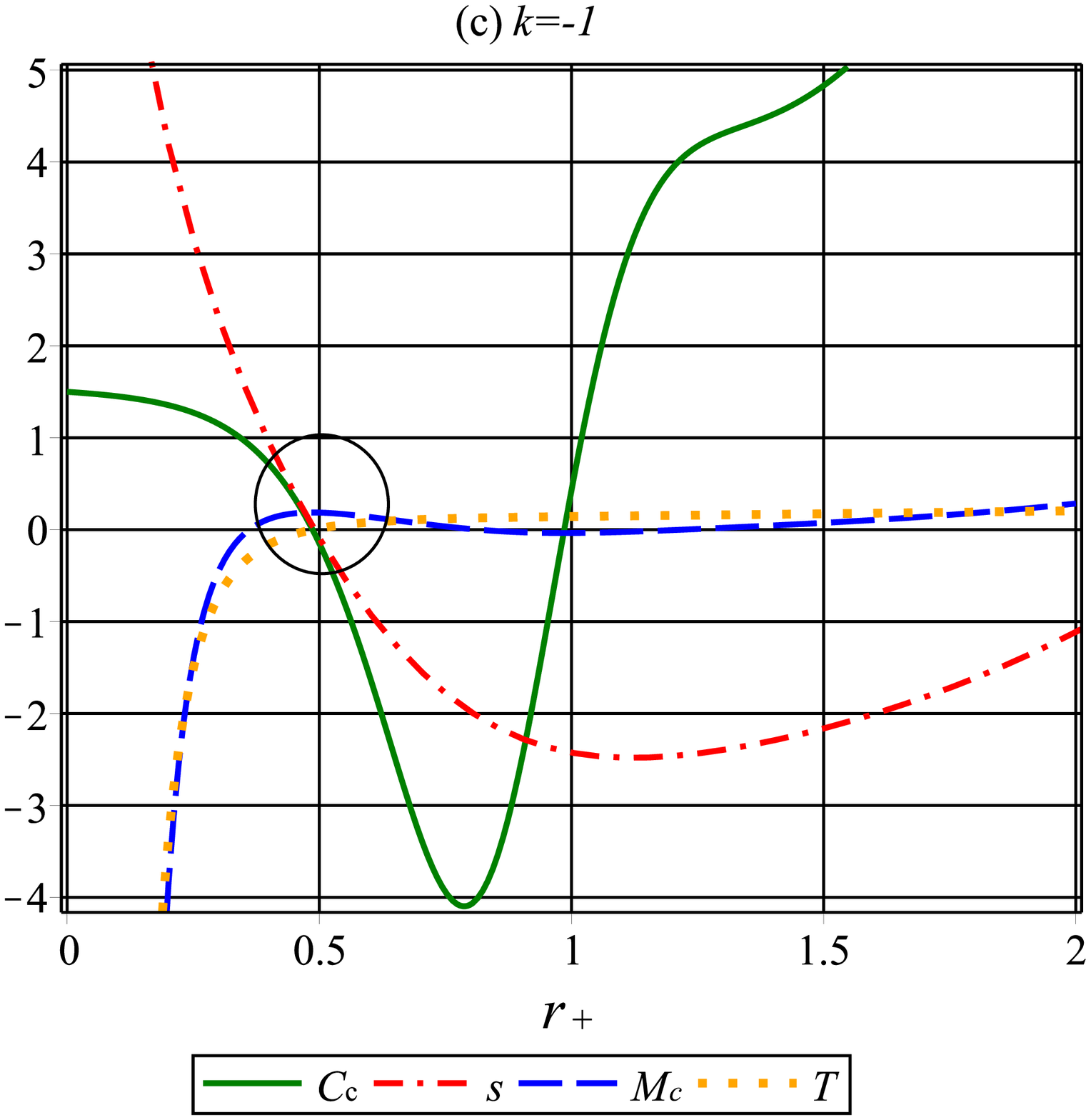} &  &  &
\end{array}%
$%
\end{center}
\caption{Logarithmic corrected entropy, mass, specific heat and temperature
in terms of $r_{+}$ for $g\left(\protect\varepsilon\right)\approx1$, $f\left(%
\protect\varepsilon\right)\approx4$, $m=1.2$ and unit value for other model
parameters.}
\label{sCM-log}
\end{figure}

We can also compute corrected specific heat as
\begin{eqnarray}  \label{CV}
C_{c}=T\left(\frac{ds}{dT}\right)_{V}.
\end{eqnarray}
Reducing the black hole size, the system goes to an unstable phase until the
entropy and specific heat vanish.\newline
Now it is interesting to study the case when black hole entropy, temperature
and specific heat are zero, but black hole mass is not zero. These are
denoted by a circle in Fig. \ref{sCM-log}. It has been shown that at a
special radius the black hole entropy, temperature and specific heat may be
zero, while black hole mass is non-zero. It is interpreted as black remnant
mass, below which the black hole will not evaporate. For $k=1$, we cannot
observe any black remnant, due to the absence of a unique (even
approximately) point where the black hole entropy, temperature and specific
heat is zero. In the case of $k=0$, there is an approximate radius $%
r_{+}\approx0.45$, where the black hole entropy, temperature and specific
heat are approximately zero, and $M_{c}\neq0$ (see Fig. \ref{sCM-log} (b)).
For $k=-1$, we find that when $r_{+}\approx0.5$, the black hole entropy,
temperature and specific heat are zero, and $M_{c}\approx0.25$ (see Fig. \ref%
{sCM-log} (c)).

%%%%%%%%%%%%%%%%%%%%%%%%%%%%%%%%%%%%%%%%%%%%%%%%

\section{Conclusion}

\label{sec6} In this paper, we have analyzed a five-dimensional black hole
solution in massive gravity coupled to the Yang-Mills theory. We have
discussed the thermodynamics of this black hole solution. We also studied
the flow of such a solution with scale, using energy-dependence of the
geometry. We have used the rainbow functions, motivated from loop quantum
gravity, hard spectrum of gamma-ray bursts and the horizon problem to
analyze such a flow with scale. It was observed that these rainbow functions
can change the behavior of the black hole thermodynamics, when the size of
black hole reduces due to the Hawking radiation.

We have also investigated the criticality in the extended phase space. It
was done by treating the cosmological constant as the dynamic pressure. Its
conjugate variable was treated as the thermodynamic volume for this black
hole solution. We have also analyzed the effects of thermal fluctuations on
this black hole solutions. It was observed that these thermal fluctuations
can be obtained from a statistical mechanical partition function for this
system. The thermal fluctuations produce a logarithmic correction for the
entropy of this black hole. We have also examined the corrections to the
specific heat for this black hole solution.

It may be noted that it would be interesting to generalize these results to
higher dimensions. Thus, we could consider higher dimensional Yang-Mills
theory coupled to massive gravity, and obtain black hole solutions in such a
theory. Then, we can analyze the thermodynamics of such solutions. Here
again, we can investigate the flow of the solution with scale, using
gravity's rainbow \cite{AK, ra12, ra14, ra18}. We can then study how such a
flow deforms the thermodynamics of such higher dimensional solutions.
Furthermore, it is expected that the thermodynamics of such solutions will
again depend on the specific rainbow functions. So, we can use rainbow
functions motivated from loop quantum gravity \cite{AC1, AC2}, the hard
spectra of gamma-ray bursts at cosmological distances \cite{AC3}, and the
horizon problem \cite{MS, h}, to deform the thermodynamics of such
solutions. It would also be pointed out to analyze the critical behavior
\cite{exten1, exten2} for this higher dimensional solutions.

We can also construct the partition function for this higher dimensional AdS
solution, and use it to analyze the thermal fluctuations for that solution
\cite{T1, T2}. It is expected that the entropy of this higher dimensional
AdS solution will again be corrected by the logarithmic correction term. It
would be useful to investigate the effects of such corrections on other
thermodynamic quantities for this higher dimensional solution. It would also
be interesting to study the effects of thermal fluctuations on the
criticality of these black hole solutions.

It may be noted that a mass term for graviton can be generated from a
gravitational Higgs mechanism \cite{higgs12, higgs14}. It would be worth to
analyze such a gravitational Higgs mechanism for various supergravity
solutions. As the Yang-Mills black hole solutions can be motivated from the
bosonic part of the low energy Heterotic string theory \cite{ymph18, ymph17}%
, it would be important to study the gravitational Higgs mechanism in low
energy Heterotic string theory. This could be used to obtain a mass term for
Yang-Mills fields. It would be interesting to investigate the consequences
of such a mass term on the thermodynamics of Yang-Mills black holes.

\end{document}